\documentclass[twocolumn,prb,preprintnumbers,amsmath,amssymb,floatfix]{revtex4}

\usepackage[dvips]{graphicx}% Include figure files
\usepackage{dcolumn}% Align table columns on decimal point
\usepackage{bm}% bold math
\usepackage{times}
\usepackage{mathptm}%\usepackage{longtable}
\def\hbar{{\mathchar'26\mkern-7muh}}
\usepackage[T1]{fontenc}
\usepackage[dvips]{color}

\begin{document}
\title{Surface energies of stoichiometric FePt and CoPt alloys and their implications for nanoparticle morphologies}
\author{Antje Dannenberg}
\author{Markus E. Gruner}
\author{Alfred Hucht}
\author{Peter Entel}
\affiliation{Faculty of Physics and Center for Nanointegration, CENIDE, University of
  Duisburg-Essen, 47048 Duisburg, Germany}
\date{\today}

\begin{abstract}
We have calculated surface energies and surface magnetic order of various low-indexed surfaces of monoatomic Fe, Co, and Pt, and binary, ordered FePt, CoPt, and MnPt using density functional theory. Our results for the binary systems indicate that elemental, Pt-covered surfaces are preferred over Fe- and Co-covered and mixed surfaces of the same orientation. The lowest energy orientation for mixed surfaces is the highly coordinated (111) surface.
We find Pt-covered (111) surfaces, which can be realized in the L1$_1$ structure only, to be lower in energy by about 400 meV/atom compared to the mixed L1$_0$ (111) surface.
We conclude that in small nanoparticles this low surface energy can stabilize the L1$_1$ structure, which is suppressed in bulk alloys. 
From the interplay of surface and bulk energies, equilibrium shapes of single-crystalline ordered nanoparticles and crossover sizes between the different orderings can be estimated.
\end{abstract}
\maketitle

\section{Introduction}
\begin{figure}
\includegraphics[width=8cm]{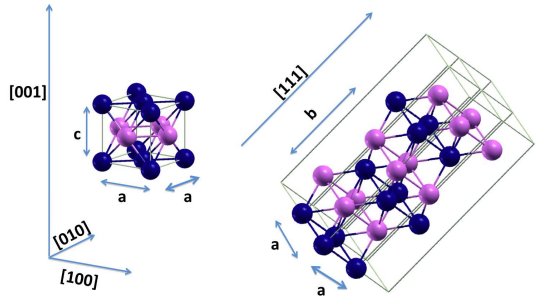}
\caption{(color online) Left: The L1$_0$-unit cell. Right: L1$_1$-cell as used in the calculations. Dark (blue) spheres denote Fe/Co/Mn atoms, light (magenta) spheres Pt atoms. In the L1$_0$ order, monoatomic planes are stacked along the [001] direction, in the L1$_1$ structure, along the [111] direction. The tetragonal distorted L1$_0$ structure has two different lattice parameters $a$ and $c$. In the L1$_1$ structure a slight distortion along the [111] direction may occur. The L1$_1$ crystal structure has only rhombohedral symmetry.}
\label{EZL10L11neu}
\end{figure}

During the last two decades, an exponential increase of the magnetic data storage areal density has been achieved. Thus, in order to continue with this trend, a constant further miniaturization of the bit size is required. Promising candidates for future ultra-high density storage media are L1$_0$-ordered FePt or CoPt nanoparticles due to their extraordinary high magnetocristalline anisotropy in the bulk phase (FePt: K$\rm_u$ = 7$\cdot10^{7}\,$erg/$\rm cm^{3}$, CoPt: K$\rm_u$ = 4.9$\cdot10^{7}\,$erg/$\rm cm^{3}$).\cite{Perez:05,Yang:04,Moser:99} The L1$_0$-lattice structure is characterized by a tetragonal distortion of a few percent along the $c$-axis accompanied by an alternating stacking of elemental layers along the [001] direction (cf.~Fig.~\ref{EZL10L11neu}). The intriguing properties of L1$_0$ FePt and CoPt alloys and nano-composites have been subject to numerous experimental and theoretical studies, e.g., see Refs. \onlinecite{Opahle:05,Ebert:09,KKRFePt:07,Shick:03,Gruner:08,Zotov:08,MacLaren:05,Podgorny:91,Mertig:95,Caroline:08,Nina:07} and references therein.\\
A subtle interplay between surface energies and internal interface energies determines the equilibrium shape of nanoparticles. One major obstacle in producing L1$_0$ FePt nanoparticles is the occurrence of multiple twinning.\cite{Olga:07,NinaPaper:07,Wang:02,Wang:08,Wmann:03,WassMannFePt:03,Fassbender:05} Multiply-twinned nanoparticles such as icosahedra or decahedra do not exhibit high uniaxial magnetocrystalline anisotropy energy due to the different crystallographic orientation of the individually ordered twins. Multiple twinning appears if the energy gain due to low surface energies exceeds the energy needed for the creation of twin boundaries. The hierarchy of surface energies is thus one important function determining the equilibrium shape of small nanoparticles.\cite{Wulff:01} Since surface energies are particularly difficult to measure in experiment, their theoretical calculation is an important task.\\
Apart from the L1$_0$ phase, also in the less common L1$_1$ structure a high uniaxial magnetic anisotropy is reported.\cite{Yamashita:97,Huang:99} In the L1$_1$ structure, alternating fcc Cu and Pt layers are stacked along the [111] direction, similar to the L1$_0$ structure, which consists of alternating (001) planes (cf.~Fig.~\ref{EZL10L11neu}). In contrast to the L1$_0$ structure, the L1$_1$ phase is only stable for bulk materials in the metallic CuPt alloy.\cite{Clark:95,Zunger:91,Bluegel:91}  Very recently, L1$_1$ type CoPt ordered films with a large magnetocristalline anisotropy, comparable in size to L1$_0$ type FePt films, were successfully fabricated.\cite{Sato:08,Hu:99} Consequently, we include also investigations of L1$_1$ ordered FePt and CoPt alloys in our study.\\

%One idea to suppress multiple twinning in small FePt-nanoparticles is to increase twin boundary energy by alloying with manganese (Mn). Furthermore the ternary alloy (Fe$_{(1-x)}$Mn$_x$)$_{50}$Pt$_{50}$ is expected to possesses an enhanced magnetocristalline anisotropy energy\cite{Thiele:06,Wills:05} which is expected to increase the stability of the L1$_0$-order. At the same time antiferromagnetism might become a problem since L1$_0$-MnPt is antiferromagetic in its ground state.
By means of density functional theory (DFT) calculations we have determined surface energies and surface magnetism of various low-index surfaces, including the (100), (001), (110), (011), and the (111) facet in the L1$_0$ phase (cf.~Fig.~\ref{SuperzellenFePtneu}) as well as the (111) surface of the L1$_1$ structure (cf.~Fig.~\ref{EZL10L11neu}).
% but there also exist earlier works\cite{Northrup:93} and recent studies\cite{Penev:05}. 
Regarding the surfaces of elemental systems, including bcc Fe surfaces\cite{Kiejna:07,Freeman:93,Weinert:83,Wang:81}, Pt (111) and (001) surfaces\cite{Fiorentini:96,Scheffler:06}, and 4d transition metal surfaces,\cite{Scheffler:91,Fiorentini:93} numerous studies can be found in literature.\cite{Rosengaard:92,Lu:05,Kiejna:05,Kiejna:99} For binary transition metal alloys however, only few investigations are available.\cite{Roesch:03,Yoo:05} \\
To the best of our knowledge, this is the first systematic first principles comparison of the energies of various low-index surfaces of FePt, CoPt and MnPt with L1$_0$  and L1$_1$ order.

\section{Method}
\begin{figure}
\includegraphics[width=7.2cm]{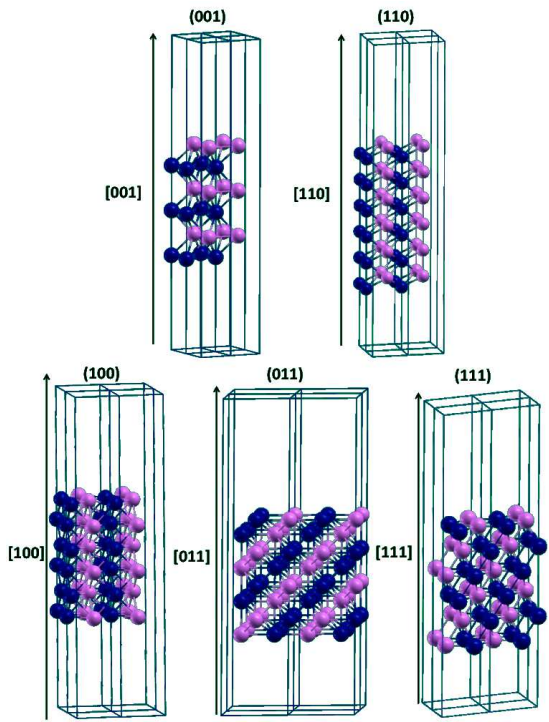}
\caption{(color online) Supercells (for clarity repeated in $x$ and $y$ -direction) used in the calculation of the surface energies in L1$_0$ FePt, CoPt and MnPt. As in Fig.~\ref{EZL10L11neu}, dark (blue) atoms are Fe/Co/Mn, light (magenta) atoms are Pt. Top left: slab for the (001) and (110) surface; bottom: the (100), (011), and the (111) surfaces. Two different surfaces appear in [001] and the [110] directions. The two surfaces are covered by different elements, while in the other cases the slabs are limited by identical mixed surfaces.}
\label{SuperzellenFePtneu}
\end{figure}

For the evaluation of the surface energies, the so called {\em slab approach}\cite {Fiorentini:96} was used. Here, the semi-infinite problem is represented by a periodically repeated two-dimensional slab with two surfaces separating the periodic images by a sufficient amount of vacuum in the third direction. Some representative slabs are shown in Fig.~\ref{SuperzellenFePtneu}.
Here it can be seen that for binary alloys two different surfaces are encountered when slabs are stacked along the [001] and [110] direction in the L1$_0$ structure and along [111] in the L1$_1$ phase. In those cases, one surface is entirely covered by Fe atoms while the other is covered with Pt atoms, respectively. Then the surface energy should be divided into two element-specific contributions as one single material component may be predominantly found at the surface. For these cases, surface-energy phase diagrams have been evaluated in order to account for the surface energies of the single material constituents. Regarding other metallic surfaces and, in specific, semiconductor surfaces, systematic investigations have been devoted to obtain structure, surface free energies, and segregation properties by, among others, the group of M. Scheffler.\cite{SchefflerHandbook:05,Scheffler:00,Moll:96,Northrup:93,Penev:05,PenevScheff:03,Kitchin:08}
%In the field of the theoretical investigation of semiconductor surfaces, great effort was made, among others, by 

\subsection{Computational details}
The self-consistent calculations are carried out with the Vienna ab initio simulation package (VASP) using a plane wave basis set and the projector augmented wave (PAW) framework.\cite{Furthmueller:96,Kresse:99} The exchange-correlation potential is used in the functional form of Perdew, Burke and Ernzerhof (PBE).\cite{Ernzerhof:96,Perdew:92} The PAW potentials include the following valence electrons: Fe: 3p$^6$3d$^7$4s$^1$, Co: 3d$^8$4s$^1$, Pt: 5d$^9$5s$^1$ and Mn: 3d$^6$4s$^1$. All plane waves with energies below the cut-off energy are included in the basis set. The cut-off energies were always chosen 25\% larger than the largest default cut-off of the element-specific potentials. We used (in eV): Fe: 366.5, Co: 335.0, Mn: 337.3, FePt: 366.5, CoPt: 335.0 and MnPt: 337.3. The integration over the Brillouin zone is done by means of finite temperature smearing (Methfessel-Paxton method for the surfaces) or tetrahedron method (for bulk systems). For the first case, the parameter $\sigma$ determines the width of the smearing (in eV): $\gamma$-Fe: 0.15, $\alpha$-Fe: 0.32, Co: 0.15, FePt: 0.2, CoPt: 0.28 and MnPt: 0.2. We used the following k-point grids: For bulk calculations: A $\Gamma$ centered (G) (13/13/13) grid. For (001) surface calculations: $\gamma$-Fe: G (19/19/1), $\alpha$-Fe: Monkhorst (M) generated (16/16/1) grid, fcc Co: M (16/16/1), Pt: G (19/19/1), FePt: M (16/16/1), CoPt: M (14/14/1), and MnPt: M (14/14/1). The parameter $\sigma$ has been carefully chosen, so that the entropy term is lower than $1 \rm meV/atom$. The electronic self-consistency iteration cycle is aborted when the energy difference between the old and the new energy is less than 10$^{-7}$ eV. The slabs used to model the surfaces consist of up to 32 atomic layers. Adjacent supercells are separated by a vacuum region of about 15 \AA~to avoid interaction between neighboring supercells (Fig.~\ref{SuperzellenFePtneu}). The geometric relaxation is done by the conjugate gradient algorithm and at least the outermost 4 layers are optimized. Relaxation was stopped when the forces were less than 0.1 eV/\AA. 
\subsection{Surface energy calculations}
\label{Calculation}

Surfaces can be created by dividing an infinite crystal into two parts. The energy needed to cut the bonds and bring the two resulting parts to infinity determines the surface energy.
% For the determination of the formation energy of a solid surface the bulk energy per atom must be known.\cite{Gross:03} 
 A straight forward procedure to calculate the surface energy is to examine the total energy $E^{\rm tot}(n)$ of a slab of the material of interest with $n$ atomic layers and to subtract $n$ times the bulk energy E$_{\rm bulk}$ of an atomic layer obtained from a separate calculation (e.g., see Ref. \onlinecite{Gross:03} for a detailed introduction):
\begin{equation} \label{simple surfEformula}
\gamma = \lim_{n \rightarrow \infty} \left \lbrace \frac{1}{2A} (E^{\rm tot}(n)-n\cdot E_{\rm bulk}) \right \rbrace.
\end{equation}
Here, $A$ is the unit surface area. For sufficiently thick slabs, bulk properties are approached in the interior of the slab and $\gamma$ is expected to converge as a function of the slab thickness towards the exact surface energy.
But proceeding as above, the surface energy rather diverges with slab thickness due to slight, unavoidable, numerical discrepancies which can be caused, e.g., by the choice of different basis or k-point grids in slab and bulk calculation.\cite{Boettger:94} In order to avoid this, the so called {\em slab approach} was proposed.\cite{Fiorentini:96} Within the slab approach, the bulk energy, $E_{\rm bulk}$, is estimated from the same slab systems for the surface energy calculations instead of using a single separate bulk calculation within a small unit cell. Therefore, the divergence problem can be avoided as consistency in all technical parameters is maintained.\\
 The quantity $E_{\rm bulk}$ is extracted as follows: First all of the total energy of various slabs with increasing thickness is calculated and plotted versus slab thickness. For large enough thicknesses the slope of a fitted straight line yields the bulk energy $E_{\rm bulk}$. In the present calculations, the surface energy converges properly if slabs with 6 atomic layers or less are discarded.\\
When an L1$_0$ FePt crystal is cleaved on any of the (100), (011), and (111) planes, the two exposed surfaces are of mixed atomic composition, i.e. they consist of the same amount of Fe and Pt atoms.
On the other hand, in the cases of both (001) and (110) cleavages, one surface consists entirely of Pt atoms while the other surface consists entirely of Fe atoms. Thus, proceeding as described above, we obtain an averaged value over the surface energies of both orientations but no information about the element specific contributions $\gamma_{\rm Fe}$ and $\gamma_{\rm Pt}$. However, a variation range for the surface energies can be given by means of surface-energy phase diagrams.\cite{SchefflerHandbook:05,Scheffler:00,Northrup:93,Moll:96} Two equivalent surfaces on top and on the bottom of the slab require to consider off-stoichiometric systems. If we compare energies of non-stoichiometric systems, the chemical potentials $\mu_i$ of the single material constituents become involved: 
\begin{equation} \label{binary surfEformula}
\gamma\{N_{\rm i}\} = \frac{1}{2A} \left( E^{\rm tot}\{N_{\rm i}\}-\sum_{i} N_{\rm i}\cdot \mu_{\rm i}\right).
\end{equation}
Here, $N_{\rm i}$ is the number of atoms of the material component $i$ and $\mu_{\rm i}$ its chemical potential. Eq.~(\ref{binary surfEformula}) is considered here only at temperature T=0. At finite temperatures, the total energy has to be replaced by the Helmholtz surface free energy. A detailed thermodynamic derivation can be found in the literature, e.g., see Refs. \onlinecite{SchefflerHandbook:05,Zangwill:88} for more details.
For the case of FePt, Eq. (\ref{binary surfEformula}) reads 
\begin{equation} \label{binary surfEformulaFePt}
\gamma(N_{\rm Fe},N_{\rm Pt}) = \frac{1}{2A} \left( E^{\rm tot} (N_{\rm Fe},N_{\rm Pt}) -N_{\rm Fe}\mu_{\rm Fe}-N_{\rm Pt}\mu_{\rm Pt}\right).
\end{equation}
The surface atoms are in equilibrium with the surrounding bulk reservoirs, which consist of the pure Fe or Pt metal and the underlying bulk alloy. Thus, the chemical potentials $\mu_{\rm Fe}$ and $\mu_{\rm Pt}$ are not independent, but related to the bulk alloy chemical potential $\mu_{\rm FePt(bulk)} = 2 \cdot E_{\rm bulk}$, the bulk chemical potentials of the elemental constituents $\mu^{\rm bulk}_{\rm Fe}$ and $\mu^{\rm bulk}_{\rm Pt}$ and the heat of the alloy formation $\Delta H_{\rm FePt}$:
\begin{equation}\label{EqNonStoich}
\mu_{\rm FePt(bulk)}=\mu_{\rm Fe} + \mu_{\rm Pt}=\mu^{\rm bulk}_{\rm Fe} + \mu^{\rm bulk}_{\rm Pt}-\Delta H_{\rm FePt}.
\end{equation}
Unlike in their bulk equilibrium phases, the chemical potentials of the single material constituents within the alloy, $\mu_{\rm Fe} + \mu_{\rm Pt}$, are not known. However, one can eliminate one of them, e.g. $\mu_{\rm Fe}$:
\begin{equation} \label{Pt surfEformulaFePt}
\gamma_{\rm Pt} = \frac{1}{2A} \left(E^{\rm tot}(N_{\rm Fe},N_{\rm Pt})-N_{\rm Fe}\mu_{\rm FePt}-\Delta N\mu_{\rm Pt}\right).
\end{equation}

Here, we assume a slightly Pt-rich environment and the surface stoichiometry is given by $\Delta$N$=\, $N$_{\rm Pt}-$N$_{\rm Fe}$ .
The stability of the bulk alloy against decomposition requests that the chemical potential $\mu_{\rm Pt}$ can take only values in the range:
\begin{equation}
\mu_{\rm Pt(bulk)}-|\Delta H_{\rm FePt}|\leq \mu_{\rm Pt} \leq \mu_{\rm Pt(bulk)}
\end{equation}
Now we can express Eq. (\ref{Pt surfEformulaFePt}) as a function of the difference in Pt chemical potential, $\mu_{\rm Pt} - \mu_{\rm Pt(bulk)}$:
\begin{eqnarray}\label{Pt surfEformulaFePtend}
\gamma_{\rm Pt} = &  \frac{1}{2A} & (E^{\rm tot}(N_{\rm Fe},N_{\rm Pt})-N_{\rm Fe} \cdot \mu_{\rm FePt(bulk)} -\Delta N\mu_{\rm Pt(bulk)}\nonumber \\
                               &                     & -\Delta N[ \mu_{\rm Pt}- \mu_{\rm Pt(bulk)}]).
\end{eqnarray}  
With the help of this equation, the so-called surface-energy phase diagrams can be determined. This approach has been applied successfully to estimate the stability of various semiconductor surface reconstructions.\cite{SchefflerHandbook:05,Penev:05,PenevScheff:03,Penev:02}

\section{Results}
\subsection{Elemental Systems}
\label{ElementalSystems}

\begin{table}
\begin{ruledtabular}
\begin{tabular}{cc cc cc cc cc}
\multicolumn{2} {c} {System}&\multicolumn{2} {c}{(111)}    &\multicolumn{2} {c}{(001)}           &\multicolumn{2} {c}{(110)} & bulk \\
                &                          &$\sigma$& M                        &$\sigma$         & M                     & $\sigma$ & M                   & M  \\ \hline
%                &                          &(eV)        &$(\mu_{\rm B})$  &          (eV)       &$(\mu_{\rm B})$&(eV)          &$(\mu_{\rm B})$& $(\mu_{\rm B})$\\
 Co (fcc)  & unr                     &0.705     &1.76                    &0.979               &1.87                   &1.398        &1.9                    & 1.63 \\
                & r                        &0.687     &1.74                     &0.964               &1.83                  &1.324         &1.85                  &       \\
 Pt (fcc)   & unr                     &0.650     &0.0                      &0.918               &0.0                     &1.370        & 0.0                    &  0.0 \\
               & r                         &0.637      &0.0                     &0.908                &0.0                    &1.305         &0.0                    &       \\
 Fe (fcc)  & unr                     &0.790     &2.71                   &0.908                &2.87                   &1.336        &2.94                   & 2.57 \\
               & r                         &0.790     &2.71                   &0.906                 & 2.86                 &1.288        &2.88                   &       \\
 Fe (bcc) & unr                     &2.434     &2.9                     &1.268                 &2.97                  &0.872        &2.6                     & 2.21 \\
               & r                         &2.355     &2.83                   &1.261                 &2.95                  &0.872        &2.6                     &       \\
\end{tabular}
\end{ruledtabular}
\caption{Surface energies, $\sigma$, in eV/atom and surface layer spin moment, M, in  $\mu_{\rm B} \rm / atom$ of the facets (111), (001), and (110) for $\alpha$- and $\gamma$-iron, fcc cobalt and platinum. Unrelaxed (unr) and relaxed (r) geometries are compared. In the last column the bulk spin moment per atom is given.}
\label{TabelleSurfEelements}
\end{table}

\begin{table*}
\begin{ruledtabular}
\begin{tabular}{l|  c c|  c c c|  c|  c c c} 
Method/                                                  &\multicolumn{2}{c} {\rm Co}     &\multicolumn{3}{c} {\rm Pt}                              &$\gamma$--Fe &\multicolumn{3}{c}{$\alpha$--Fe}                    \\ 
Source                                                   &   (111)           &   (001)            & (111)             & (001)              & (110)             &              (111)   & (111)             & (001)               & (110)             \\ \hline
DFT$^{a}$                                              &0.687 (2.045)&0.964 (2.110)   &0.637 (1.490) &0.908 (1.840) &1.305 (1.869) &0.790 (2.203)   &2.355 (2.694) &1.261 (2.499)    &0.872 (2.444)\\ \hline 
DFT\cite{Iddir:07}$^{,b}$                      &                      &                       &0.660 (1.535)&0.915 (1.843) &1.308 (1.863)  &                         &                      &                          &                       \\
DFT\cite{Getman:07}$^{,b}$                &                      &                       &0.620 (1.450) &                        &                      &                          &                      &                        &                      \\
DFT\cite{Blonski:04}                             &                      &                       &                      &                       &                       &                         &2.220 (2.540) &1.135 (2.250)  &0.803 (2.250)\\
DFT\cite{Spencer:02}$^{,b}$                &                      &                       &                       &                      &                       &                        &2.203 (2.520) &1.155 (2.290)   &0.810 (2.270) \\
DFT\cite{Vitos:98}$^{,c}$                          &                      &                       &                      &                      &                      &                         &2.694 (2.733) &1.265 (2.222)    &0.978 (2.430)  \\
DFT\cite{Freeman:93}$^{,k}$                    &                      &                       &                      &                      &                      &                         &2.972 (3.400) &                         &                        \\
DFT\cite{Scheffler:06}$^{,i}$                     &                      &                       &0.710 (1.661)&                      &                      &                         &                      &                             &                         \\
DFT\cite{Scheffler:06}$^{,j}$                     &                      &                       &0.610 (1.427)&                      &                      &                         &                      &                             &                         \\
DFT\cite{Kiejna:07}$^{,e}$                        &                      &                       &                       &                      &                     &                        &2.260 (2.580) &1.250 (2.470)      &0.850 (2.370) \\
DFT\cite{Fiorentini:96}$^{,d}$                    &                       &                       &                       &1.245 (2.522) &                     &                       &                       &                            &                        \\ 
DFT\cite{Alden:92}$^{,f}$                          &0.907 (2.700)   &1.270 (2.780)&                        &                      &                     &                       &                      &1.100 (2.180)     &0.949 (2.660)    \\ 
DFT\cite{Rosengaard:92}$^{,g}$               &1.100 (3.230)    &                     &0.980 (2.350) &1.190 (2.480)   &                      &1.150 (3.280)   &                     &                           &1.120 (3.090)     \\    
TB\cite{Papaconstantopoulos:96}$^{,h}$&                         &                       &1.073  (2.510) &1.397 (2.830) &2.074 (2.970)&                         &                      &                             &                      \\ 
MEAM\cite{Baskes86:86}$^{,l}$               &                        &                       &0.616 (1.440) &0.814 (1.650)    &1.222 (1.750)&                        &                      &                              &                     \\ 
MEAM\cite{Baskes:92}$^{,l}$                    &                        &                       &0.710 (1.660) &1.071 (2.170)    &1.487 (2.130)&                         &1.503 (1.720) &1.155 (2.289)      &0.559 (1.566)\\ 
Exp.\cite{Miller:77}$^{,m}$                       &                        &                       &                      &($2.490_{\rm av}$)&                     &                          &                      &($2.360_{\rm av}$) &                   \\ 
Exp.\cite{Boer:89}$^{,n}$                          &($2.550_{\rm av}$)&                    &                      &($2.480_{\rm av}$)&                      &                          &                      &($2.475_{\rm av}$)&                   \\
Exp.\cite{Tyson:75}$^{,m}$                       &                          &                       &                      &($2.370_{\rm av}$) &                    &($2.170_{\rm }$)  &                      &                           &                   \\
\end{tabular}
\end{ruledtabular}
\caption{Summary of relaxed surface energies in $\rm eV/atom$ ($\rm J/m^2$) for fcc Co, Pt, fcc Fe, and bcc Fe as calculated by the authors in comparison to data taken from the literature.\\
$^{a}$ Authors: GGA (PBE), VASP. \\
$^{b}$ GGA (PW91), VASP. \\
$^{c}$ Full charge density (FCD) method in the GGA, based on linear muffin tin orbitals in tight binding (TB-LMTO) and atomic-sphere approximation (ASA).
 Unrelaxed surface energies, using a lattice constant $\rm a = 3.001$\AA.\\
$^{d}$ LDA, FP-LMTO, and Ceperly-Alder parametrization for xc-potential, slab approach.\\
$^{e}$ VASP, TB-LMTO spinpolarized GGA for xc-potential.\\
$^{f}$ TB-LMTO, ASA, Ceperly-Alder for xc-potential.\\
$^{g}$ TB-LMTO, ASA, Ceperly-Alder for xc-potential. Surface relaxation neglected. Using experimental lattice constant.\\
$^{h}$ Tight binding method (TB). Unrelaxed surfaces geometries.\\
$^{i}$ All-electron full-potential linearized augmented plane-wave FPLAPW (WIEN97), GGA(PBE).\\
$^{j}$ FPLAPW (WIEN97), GGA(PBE). Using the experimental lattice constant.\\
$^{k}$ FPLAPW, LDA, Barth and Hedin formula for exchange-correlation potential.\\
$^{l}$ Empirical, modified embedded-atom method: MEAM.\\ 
$^{m}$ Extrapolation from experimental solid-vapour surface energies at higher temperatures to $T=0$ K (approximation for an "averaged (av)" polycrystalline surface). The solid-vapour surface energy is derived from liquid surface-tension measurements.\\
$^{n}$ Estimation of surface energy by subtracting from the measured surface-tension of the liquid an entropy term propotional to the melting temperature.}
\label{LiteratureElements}
\end{table*}

For the elementary systems, the properties of the bulk phases are well known and have been reported previously for most cases.\cite{Herper:99,EntelIron:00,WassermannIron:94,Rosato:93,Khein:95,Kresse:97,Ptbulk:99,Cobulk:99}
Our results match well with these investigations especially with those, where similar methods and technical parameters were used.\cite{Kresse:97,EntelIron:00,Ptbulk:99,Cobulk:99}
For metals with fcc lattice structure, we could confirm the following trend: With decreasing coordination number of the surface atoms, the surface energy, $\sigma$, and the spin moment, M, of the outermost surface layer increases. This correlation is well known in literature\cite{Gross:03,Scheffler:91} and leads for fcc metals to: 
$$
\begin{array}{rlrlr}
\sigma(111) & < &  \sigma(001) & < &  \sigma(110),\\
\rm M(111) & < &  \rm M(001) & < &  \rm M(110).\\
%\Delta \sigma(111) & < &  \Delta \sigma(001) & < &  \Delta \sigma(110) \\
\end{array}
$$
These results match intuition since the surface energy (the energy needed for cutting some "bonds") must grow with decreasing surface coordination number.
The coordination numbers ($z$) for the fcc structure are: $z = 7$ for the most open surface (110), $z = 8$ for the (001) facet and $z = 9$ for the most densely packed (111) surface orientation. For the very open bcc geometry we have: $z = 4$ for the (001) and the (111) surface and $z = 6$ for the (110) surface.
Table \ref{TabelleSurfEelements} presents our calculation of surface energies, $\sigma$, in eV/atom and magnetic moment on surface atom, M, in $\mu_B$/atom of all considered elemental systems. Results for unrelaxed (unr) as well as relaxed (r) structures are presented. In the last column, the bulk magnetic moment is given. 
For platinum, a large variation in surface energy (by a factor of two) between the most open (110) surface and the densely packed (111) surface is found. In the surface layers, the spin moment is enhanced between $2$ and $8$\% compared to the bulk value. This is in accordance with the observation of increased magnetic moments in low dimensional systems, e.g., as in small Fe clusters.\cite{Sahoo:04,Sahoo:06}\\
 In Table \ref{LiteratureElements}, our results for the surface energies are compared to data available in the literature. 
 For Pt, we find very good consistency of our surface energies with the {\em ab initio} calculations of Ref.\,\onlinecite{Getman:07}, \onlinecite{Scheffler:06}, \onlinecite{Iddir:07}. Concerning the work of da Silva {\em et al.}\cite{Scheffler:06} the deviations remain in the range of 4-8\%. They use the same functional form for the exchange correlation potential (PBE) and their slabs are relaxed as well. Noteworthy deviations occur in comparison with Ref.\,\onlinecite{Fiorentini:96}, where DFT calculations within the local density approximation (LDA) are performed. The authors use seven layer of vacuum in between adjacent supercells and do not relax their slabs. Their surface energy for the Pt (001) facet $\rm \sigma_{\rm Pt}(001)=1.245$~eV/atom is by 27\% larger than our value of $\sigma_{\rm Pt}(001)=0.908$~eV/atom, and presumably related to the different choice of the exchange-correlation potential. For 3$d$-transition metals the LDA is known to show strong over-binding, i.e. cohesive energies turn out to be too large and lattice constants too small compared to experiment. As the surface energy is correlated to the cohesive energy, the values obtained within the LDA can be expected to be too large as well. This is different for the gradient corrected exchange correlation potential where a slight overestimation of the lattice constant is common.\\
 The values reported by Skriver {\em et al}\cite{Rosengaard:92} using the tight-binding linear-muffin-tin orbital approach (TB-LMTO) with the atomic sphere approximation (ASA) are considerably larger than our PBE-values for all systems under investigation, e.g., larger by 0.4 eV/atom for Co (111). Similar considerations hold true for the data of Ref.\,\onlinecite{Papaconstantopoulos:96} which again show strong deviations to larger values. For completeness, we list in Table \ref{LiteratureElements} also semi-empirical and empirical methods, as e.g. the tight-binding and the modified embedded atom method (MEAM). These approaches do not yield the same accuracy of DFT methods but are frequently used for large scale simulations. Here, free parameters are fitted to reproduce certain surface properties and thus can no longer be considered as high-level {\em ab initio} investigations.\\
  Also for $\alpha$--Fe very good agreement with other first-principle DFT calculations (Ref.\,\onlinecite{Blonski:04}, \onlinecite{Spencer:02}, \onlinecite{Kiejna:07}) is achieved. Slight deviations occur compared to the results of Ref.\,\onlinecite{Vitos:98}, \onlinecite{Freeman:93}, \onlinecite{Alden:92} which may again be in part related to the different approximations for the exchange correlation potential as discussed above, different lattice constants or missing relaxation of the surface layer.\\
   If experimental data are available at all, they mostly are obtained by liquid metal surface-tension measurement at higher temperatures and are then extrapolated to $T=0$~ K. Thus, they are "averaged" values which can not be attributed to a special surface orientation.\\
  The large discrepancies found in literature show that the calculation of surface energies is a rather delicate task. They should be interpreted in terms of a comparison to other values. This underlines the necessity for a systematic and (technically) consistent comparative investigation of single element and binary transition metal surfaces as presented in this work.

\subsection{Binary Alloys of Fe, Co, Mn with Pt}

\begin{figure*}
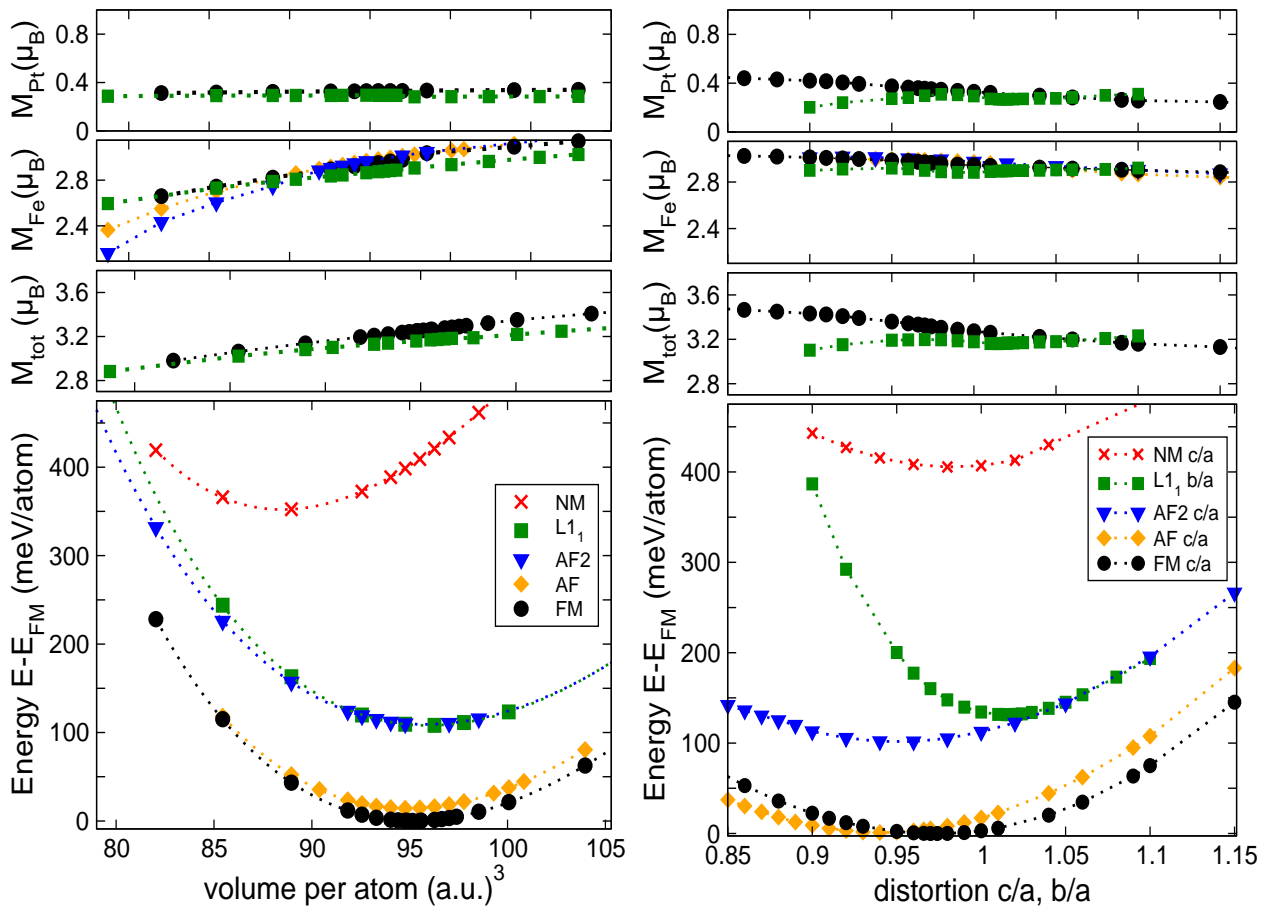

\includegraphics[width=8.2cm,height=12cm,clip]{GFePtallPAPNEU.eps}\hspace{0.2 cm}\includegraphics[width=8.2cm,height=12cm,clip]{GFePtcabaPAPNEU.eps}\\
\caption{(color online) Left: Energy versus volume curves and element-specific magnetic moment of L1$_0$ FePt for different magnetic structures and for the L1$_1$ phase. The ferromagnetic phase (FM, black circles) leads to the equilibrium structure. The antiferromagnetic phase (AF, orange diamonds) appears only 13.6 meV/atom higher in energy and shows the competition between ferro- and antiferromagnetism. The ordered L1$_1$ structure (L1$_1$, green squares), the MnPt-type antiferromagnetic structure (AF2, blue downward triangles), and the non-magnetic phase (NM, red crosses) are not stable for bulk FePt.
Right: Energy (distortion) with fixed volume for L1$_0$ and L1$_1$ FePt and different magnetic states. The ferromagnetic phase minimizes the energy at a $c/a$-ratio of 0.974 while the antiferromagnetic structure becomes stable at a slightly smaller $c/a$-ratio. For the L1$_1$ structure the $b/a$-ratio is varied and the energy minimum is found at $b/a = 1.015$.
In the upper panels the magnetic moments M$_{\rm Fe}$ and M$_{\rm Pt}$ (in $\mu_{B}$ per atom) are given as well as their sum M$_{\rm tot}$ (in $\mu_{B}$ per cell).}
\label{GFePtallNORMPaper}
\end{figure*}

\begin{figure*}
\includegraphics[width=8.2cm]{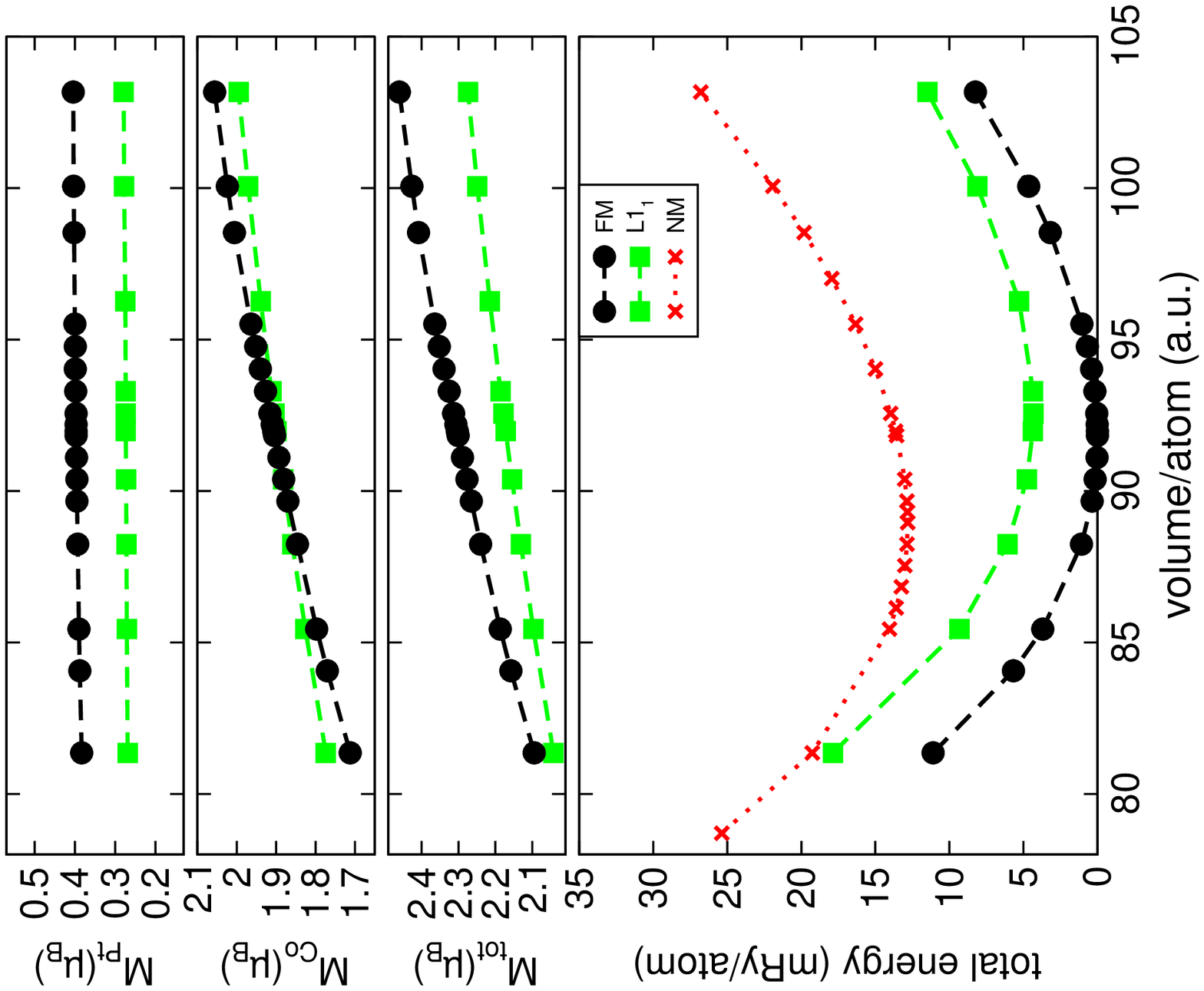}\hspace{0.2cm}\includegraphics[width=8.2cm]{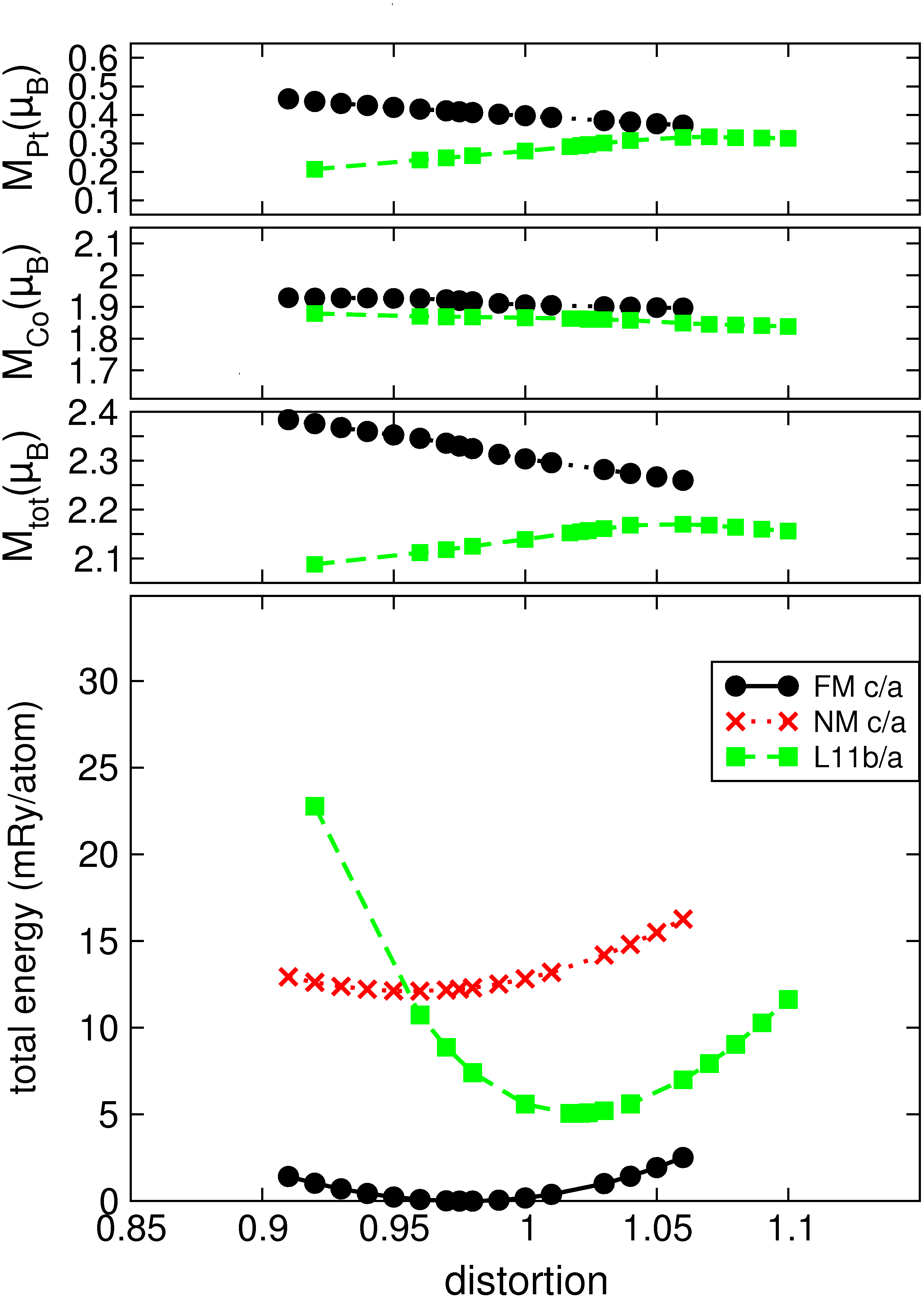}\\
\caption{(color online) Left: L1$_0$ CoPt: As for FePt, the ferromagnetic phase (FM, black circles) leads to the equilibrium structure of L1$_0$ CoPt. The ordered L1$_1$ structure (L1$_1$, green squares) appears at roughly 68 meV/atom higher in energy.
Right: For L1$_0$ CoPt, energy is minimal for the ferromagnetic phase at a $c/a$-ratio of 0.976. The L1$_1$ structure has minimum energy for $b/a = 1.017$.}
\label{GCoPtcabaAllNORMPaper}
\end{figure*}

%\begin{figure}
%\includegraphics[width=8.2cm]{GCoPtallNORM.eps}
%\caption{(color online) L1$_0$ CoPt: As for FePt, the ferromagnetic phase (FM, black circles) leads to the equilibrium structure of L1$_0$ CoPt. The ordered L1$_1$ structure (L1$_1$, green squares) appears at roughly 5 mRy/atom higher in energy.}
%\label{GCoPtallNORM}
%\end{figure}

%\begin{figure}
%\includegraphics[width=8.2cm]{GCoPtcabaAllNORMDiplom.eps}
%\caption{(color online) For L1$_0$ CoPt, energy is minimal for the ferromagnetic phase at a $c/a$-ratio of 0.976. The L1$_1$ structure has minimal energy for $b/a = 1.017$. Symbols as in Fig. \ref{GCoPtcabaAllNORMPaper}.}
%\label{GCoPtcabaAllNORMDiplom}
%\end{figure}

The structural and energetic properties of bulk Pt-based alloys with L1$_0$ order have also been subject to numerous theoretical and experimental surveys.\cite{Wills:05,Shick:03,Eriksson:01,Galanakis:00,Garg:99,Zotov:08,MacLaren:05,Mertig:95}
However, as we also deal with the less well studied L1$_1$ structure, we provide a detailed comparison of both phases in the following.
\subsubsection{L1$_0$ and L1$_1$ bulk phases of FePt and CoPt}

In the left panel of Fig.~\ref{GFePtallNORMPaper}, the energies of different magnetic structures of the ordered L1$_0$ and L1$_1$ phases are compared as a function of atomic volume.
The ferromagnetic (FM) phase with L1$_0$ order is the ground state with a lattice constant of $3.835$\,\AA. The layer-wise antiferromagnetic order (AF) is only 13.6 meV/atom higher in energy. This shows the competing behavior between ferromagnetism and antiferromagnetism, which has been predicted previously from {\em ab initio} calculations for L1$_0$ FePt.\cite{Zeng:02,Brown:03,MacLaren:05} The L1$_1$ structure is characterized by an equilibrium lattice parameter of $a = 3.844$\,\AA \,and is found to be 122 meV/atom higher in energy. This is in agreement with the experimental observation that bulk L1$_1$ FePt is not stable. However, the energy difference between the phases decreases as the valence electron concentration increases.\cite{MyICML11:09}
As expected, the total magnetic moment, $\rm M_{\rm tot}$, in the FM L1$_0$ structure and FM L1$_1$ structure is dominated by the Fe spin moment, $\rm M_{\rm Fe}$, and steadily increases with increasing volume. The induced Pt moment, $\rm M_{\rm Pt}$, follows the trend of the Fe moments.
The optimum $c/a$-ratio is determined keeping the volume at the energetic minimum of the cubic structures fixed (cf.~Fig.~\ref{GFePtallNORMPaper}, right).   
For the ferromagnetic phase, $c/a = 0.974$ minimizes the total energy while the antiferromagnetic phase becomes stable at a slightly lower $c/a$-ratio. For the L1$_1$ structure the $b/a$-ratio is varied and shows that $b/a = 1.015$ minimizes the total energy (cf.~Fig.~\ref{GFePtallNORMPaper}).\\
The total magnetic moment in the L1$_0$ structure decreases with increasing $c/a$-ratio, while in the L1$_1$ phase shows only little variation. 
Analogous bulk calculations have been also carried out for CoPt and MnPt. MnPt possesses an antiferromagnetic groundstate with $a = 3.887$\,\AA~and $c/a = 0.937$. For bulk CoPt, the energy versus volume curve and the $c/a$ -variation (and $b/a$ -variation for the L1$_1$ structure) is shown in Fig.~\ref{GCoPtcabaAllNORMPaper}. We find the FM L1$_0$ structure to be the most stable one with an equilibrium lattice constant of $a = 3.793$\,\AA~and $c/a = 0.976$. In the FM L1$_1$ structure a lattice parameter of $a = 3.801$\,\AA~and a $b/a$-ratio of 1.017 is obtained. The FM L1$_1$ structure is $68\, {\rm meV/atom}$ higher in energy.
 The magnetic spin moments show qualitatively the same behaviour as for FePt. But interestingly, the induced Pt moment is as high as in the case of FePt, even though the spin moment of the Co atom is clearly lower than the spin moment of the Fe atom. Thus, the hybridization between the Co and the Pt $d$-electrons seems to be stronger than in the FePt alloy. Furthermore, for CoPt, the hybridization is stronger in the L1$_0$ phase than in the L1$_1$ phase.\\

\subsubsection{Surface properties of FePt, CoPt, and MnPt}\label{SurfFePt}

\begin{figure}
\includegraphics[width=8.2cm]{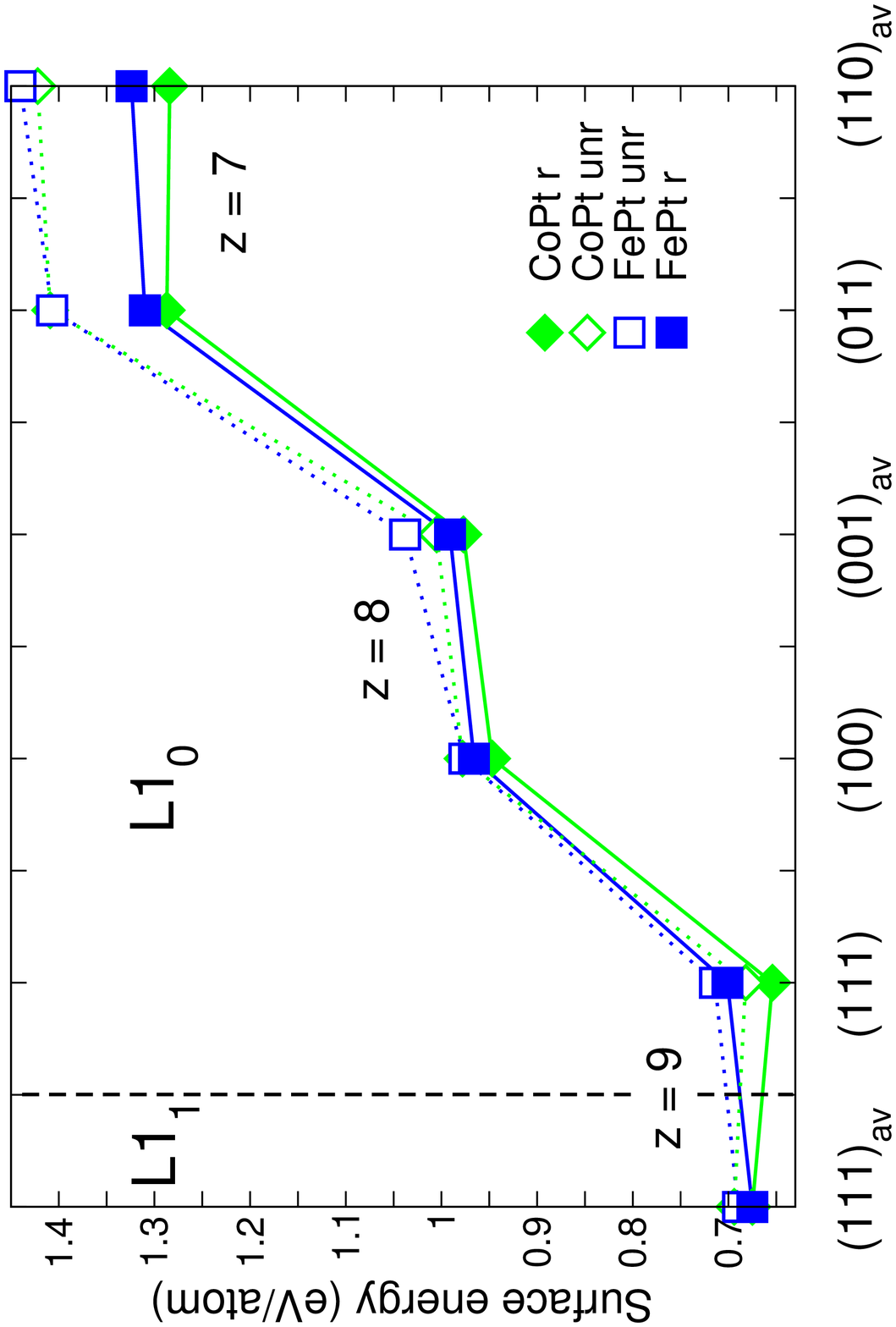}
\caption{(color online) FePt (blue squares) and CoPt (green diamonds) surface energies for the (111), (100), (001), (011), (110) facets in L1$_0$, and the (111) facet in L1$_1$ structure. The corresponding coordination numbers, $z$, are given as well. Unrelaxed (open symbols) as well as relaxed values (filled symbols) are shown. For the special orientations with two different surfaces, averaged (av) values over both surface energies are given. This applies in L1$_0$ for (001) and (110), and in L1$_1$ for (111). The highly coordinated (111) surfaces in L1$_1$ phase and L1$_0$ structure are energetically clearly favored. The lines are only guides to the eye.}
\label{GSurKNFePtCoPtRLXpaper}
\end{figure}

\setlength{\tabcolsep}{4pt}
\begin{table*}
\begin{ruledtabular}
\begin{tabular}{cc| ccc| ccc| cc}
\multicolumn{2} {c} {(hkl)/Structure}&\multicolumn{3} {c}{FePt (fct)}                                                 &\multicolumn{3} {c}{CoPt (fct)}                                               &\multicolumn{2} {c}{MnPt (fct)}           \\
                                         &             &\multicolumn{3} {c}{FM,\,$\rm M_{bulk}=1.63\,(\mu_B/atom)$} &\multicolumn{3} {c}{FM,\,$\rm M_{bulk}=1.14\,(\mu_B/atom)$} &\multicolumn{2} {c}{AF,\,$\rm |M_{Mn}|=3.58\,(\mu_B/atom)$} \\
                                        &              &$\sigma$&$\gamma$        &M                                                &$\sigma$  & $\gamma$    & M                                                & $\sigma$               &M        \\ 
                                        &              &(eV/atom) &$\rm (J/m^2)$&$\rm (\mu_B/atom)$                   &(eV/atom)&$\rm (J/m^2)$&$\rm (\mu_B/atom)$                    &(eV/atom) &$\rm (\mu_B/atom)$\\ \hline
 (111)$_{\rm av}$/L1$_1$&unr          &0.690         &                       & 1.63         	                             &0.694       &                      &1.15                                    	         &                           	 &            \\
                                         &r           &0.675          &1.781              & 1.63                                          &0.675       &1.717              &1.15                                   	          &                               &              \\ \hline                          
 (100)/L1$_0$                   &unr         &0.976         &                        &1.78                                           &0.978        &                      &1.25                                   	           &                             &          \\
     					&r	     &0.967          &2.125               &1.75                                           &0.947        & 2.125            &1.23                      	                    &                            &           \\
 (011)/L1$_0$                   &unr         &1.407          &                        &1.82                                           &1.409        &                      &1.26                      	                    &                           &           \\
 					&r           &1.310           &2.008            &1.81                                            &1.287        & 2.024           &1.26                        	                     &                            &         \\  
 (111)/L1$_0$                   &unr         & 0.714            &                     &1.71                                           & 0.682        &                    &1.19                                                 &0.649                 &3.80    \\
 					&r	     & 0.701             &1.763            &1.69                                           & 0.654       &1.680             &1.19                                               &0.626                 &3.83         \\
 (001)$_{\rm av}$/L1$_0$ &unr        &1.038              &                      &1.70                                          &1.005         &                     &1.19                      	                     &1.025                   &3.89     \\
 					&r	     &0.991             &2.121            &1.70                                            &0.977          & 2.192           &1.20                                              &0.986                   &3.85     \\          
 (110)$_{\rm av}$/L1$_0$  &unr       &1.440             &                      &1.80                                             &1.422         &                     &1.27                                             &                               &               \\
 					&r          &1.342              &2.085            &1.77                                             &1.284         &2.039            &1.57                                              &                               &              \\
\end{tabular}
\end{ruledtabular}
\caption{Surface energies, $\sigma$, of various low-indexed facets in eV/atom and magnetic moment on surface atom, M, in $\mu_{\rm B}$/atom for FePt, CoPt, and MnPt. The $c/a$-ratios of the considered face centered tetragonal structures (fct) are: FePt: 0.974, CoPt: 0.976, MnPt: 0.937. Values are given for unrelaxed (unr) as well as relaxed (r) surfaces. For the (001) and the (110) surfaces of the L1$_0$ phase and the (111) facet of the L1$_1$ phase, only averaged (av) values over both terminations can be given. For the antiferromagnetic MnPt alloy the absolute value of the Mn atom, $\rm |M_{Mn}|$ is shown.
For L1$_0$ FePt an excellent agreement with the results of Ref. \onlinecite{Yoo:05} is obtained (deviations of less than 3\%).}
\label{TabelleSurfEbinary}
\end{table*}

\begin{figure}
\includegraphics[width=8.2cm]{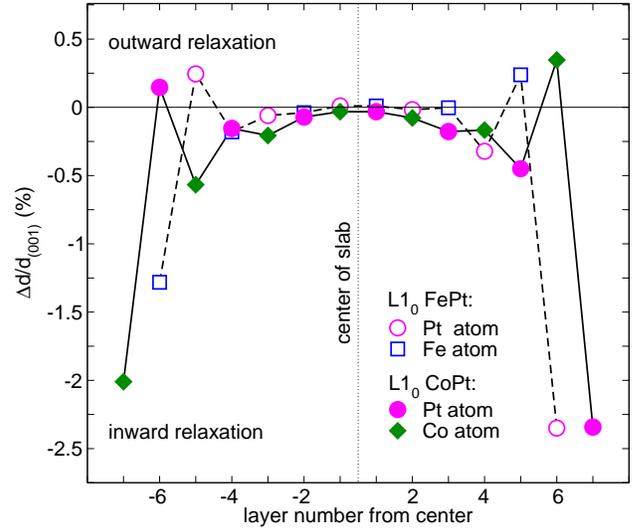}
\caption{(color online) Relaxed atomic positions of a 12-layer thick FePt and 14-layer thick CoPt (001) slab in terms of the relative displacement $\Delta {\rm d/d_{(001)}}$. The outermost layers move inwards by a few percent in all cases. After the fifth layer beneath the surface of FePt (sixth layer for CoPt), the oscillatory expansion and compression of the ideal bulk lattice parameter disappears. Open (blue) squares denote Fe atoms, filled (green) diamonds Co atoms. Circles (magenta) belong to Pt atoms, in the FePt alloy as open symbols, in the CoPt alloy as filled symbols.}
\label{GCoPtFePtRlxPosDIPLOM}
\end{figure}

Most preceding surface energy studies for binary alloys have been carried out for ideal cleaved surfaces, neglecting the effects of possible relaxations. To close this gap and to give an account on the importance of relaxations in FePt and CoPt surfaces, the surface energies of L1$_0$ and L1$_1$ FePt and CoPt with subsequent relaxation of the atomic positions are compared in Fig.~\ref{GSurKNFePtCoPtRLXpaper} for all investigated surfaces.
 For the slabs with two different surfaces the values are averaged over both possible terminations, i.e., for the L1$_0$ phase, the (001) and the (110) facet and in L1$_1$ the (111) facet.\\
Lowest surface energies are found for the highly coordinated (111) facet: In the case of L1$_1$ FePt and CoPt we find: $\sigma (111) = 0.675$\,eV/atom, in the case of L1$_0$ FePt: $\sigma (111) = 0.701$\,eV/atom, and for L1$_0$ CoPt our calculations yield: $\sigma = 0.654$\,eV/atom (cf.~Table~\ref{TabelleSurfEbinary}). These results may to some extent explain the trend that in gas phase experiments frequently FePt icosahedral nanoparticles with platinum covered (111) surfaces are generated.\cite{Wmann:03,WassMannFePt:03,Fassbender:05} The more open (100) facets, which occur in the L1$_0$ cuboctahedron (cf.~Fig.~\ref{FePtL10WP}) which is desired for magnetic data storage media, lie with $\overline{\sigma}(001)\simeq 1 {\rm eV/atom}$ higher in energy. The highest surface energies are found for the most open facets (011) and (110). Qualitatively similar results have been found for L1$_0$ PdZn and PtZn.\cite{Roesch:03} Again, the surface energy decreases with increasing coordination numbers analogous to the trend for the fcc metals.
The modification of the surface energy by relaxation is maximum for the most open surfaces (cf.~Fig.~\ref{GSurKNFePtCoPtRLXpaper}). Here, a reduction of the surface energy of about\,7\% occurs while for the densely packed (111) facet the reduction amounts only to approximately\,2\%.
 %The layer resolved atomic relaxation processes is shown in Fig.\,\ref{GCoPtFePtRlxPosDIPLOM}.
 The layer resolved atomic relaxation processes is shown in Fig.~\ref{GCoPtFePtRlxPosDIPLOM}.
The outermost layers of the slab move slightly inwards. Here, the relative displacement amounts to about 1.25\% for the Fe surface layer, about 2\% for the Co surface layer, and about 2.3\% for the Pt surface layers. In the interior of the slab, the relaxation shows an oscillating behaviour which disappears for FePt beneath the fourth subsurface layer. This agrees with previous findings for bcc Fe surfaces\cite{Kiejna:07} and FePt nanoparticles\cite{Gruner:08}.
 The relaxation in this (001) slab system is significantly lower than for the (111) facet of a L1$_0$ FePt cuboctahedron (ca. 8\%).\cite{Nina:07,NinaPaper:07} This agrees well with the experimental findings in the case of cuboctahedra, where no noteworthy relaxation of the (001) and (110) surfaces is found.\\ 
 The relaxation behaviour of elementary metal surfaces has been subject to various studies in the past six decades.\cite{Smoluchowski:41,Finnis:74,Pettifor:78,Hill:80,Heine:86,Scheffler:91,Zolyomi:09}
 For transition metals, it has been attributed to the competing influence of the partial pressures arising from the localized $d$-bonds on the one side and the $sp$-electrons on the other, which are partially relieved at the surface.\cite{Pettifor:78} However, the picture for the complete transition metal series is not uniform. While for most transition metal systems with a nearly half-filled $d$-band strong inward relaxation is observed, the effect diminishes towards the end of the series and eventually reverses sign for the $5d$ noble metals Pt and Au.\cite{Heine:86,Scheffler:91,Zolyomi:09} The effective inwards relaxation, which we observe for the binary FePt and CoPt surfaces is thus certainly influenced by the hybridization of the $3d$ and $5d$ electrons within the surface and subsurface layer.
%The multilayer relaxation for transition metals can be qualitatively explained in the framework of an electrostatic model.\cite{Scheffler:91,Hill:80} As layed out by Pettifor\cite{Pettifor:78}, the equilibrium lattice constant is determined by the competition between the homogeneous outward pressure of the $sp$-electrons and the inward force of localized $d$-bonds. At the surface the atoms loose some of the binding partners. Here, the outward pressure of the $sp$-electrons can escape into the vacuum ("spill out"), while the $d$-bonds between the first two layers remain essentially unchanged. Furthermore the $s$-electrons smear out in order to reduce their kinetic energy and to avoid charge accumulation. What remains is the inwards force of the localized $d$-bonds, which causes the inwards relaxation of the outermost atoms.
\par
Now we turn to the question how the single material constituents contribute to the averaged values. The exact determination of, for example, $\sigma_{\rm Pt}$(001) in FePt (cf.~Fig.~\ref{SuperzellenFePtneu}) is not possible within the slab approach due to the missing knowledge of the chemical potential of the material components in the alloy. But the range of variation for the surface energy can be given by means of surface-energy phase diagrams. Here the two limiting cases for $\sigma_{\rm Pt}$(001) in FePt are calculated by artificially varying the stoichiometry and with this the difference in Pt chemical potential $\Delta \mu_{\rm Pt} = \mu_{\rm Pt}-\mu_{\rm Pt(bulk)}$ as explained in section \ref{Calculation}, cf.~Eq.~(\ref{Pt surfEformulaFePt}).\\
\begin{figure*}
\includegraphics[width=12.5cm]{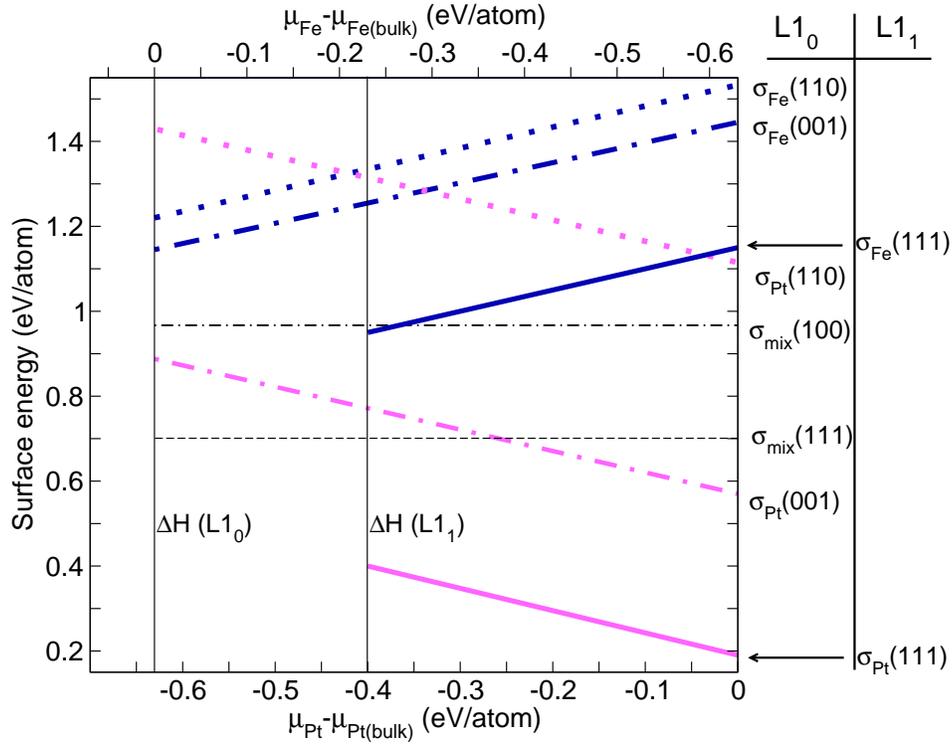}
\caption{(color online) Equilibrium surface-energy phase diagram of L1$_0$ and L1$_1$ FePt including all investigated surface orientations. On the lower horizontal axis the difference between the platinum chemical potential in the bulk phase, $\mu_{\rm Pt(bulk)}$, and in the alloy FePt, $\mu_{\rm Pt}$, is given (for the case of iron on the upper horizontal axis, respectively). The maximum difference is given by the formation enthalpy, $\Delta$H$_{\rm FePt}^{\rm L1_0}= -0.6305$ eV/atom in the L1$_0$ phase and $\Delta$H$_{\rm FePt}^{\rm L1_1}= -0.4$ eV/atom in the L1$_1$ phase, and symbolized by the two vertical lines. Thick bright (magenta) lines denote Pt covered surfaces, corresponding thick dark (blue) ones Fe covered ones. In the L1$_0$ structure we use dotted lines for the most open (110) facet and dot-dashed lines for the (001) facet. The mixed (100) and (111) facets are shown as thin horizontal (black) lines. The (111) surfaces associated with the L1$_1$ bulk ordering are denoted by thick solid lines.}
\label{GsurfDiagrFePtallDRESDENpaper}
\end{figure*}
\begin{figure*}
\includegraphics[width=12.5cm]{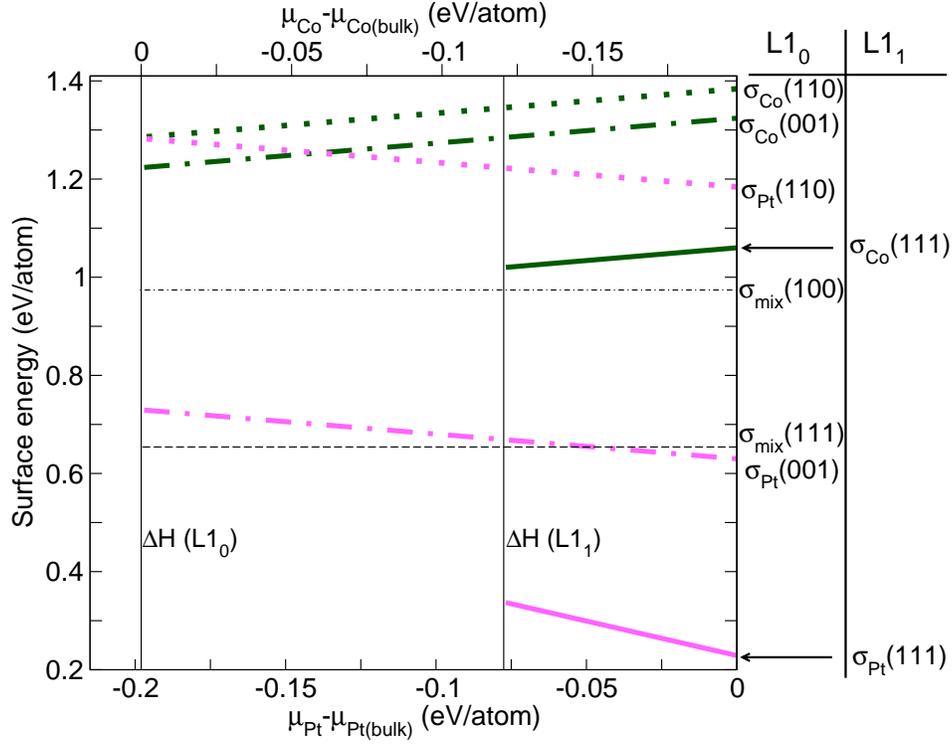}
\caption{(color online) Equilibrium surface-energy phase diagram of L1$_0$ and L1$_1$ CoPt for the (100), (001), (110), and the (111) surface orientations. Vertical lines mark the formation enthalpy $\Delta$H$_{\rm CoPt}^{\rm L1_0}= -0.198$ eV/atom in the L1$_0$ phase and $\Delta$H$_{\rm CoPt}^{\rm L1_1}= -0.076$ eV/atom in the L1$_1$ phase. Line assignment and color coding as in Fig.~\ref{GsurfDiagrFePtallDRESDENpaper} except for Co covered surfaces (thick lines, dark green). As already found for FePt the Pt covered surfaces are energetically preferred over Co covered or mixed surfaces of the same orientation.}
\label{GSurfDiagrCoPtallPAPER}
\end{figure*}
\begin{figure*}
\includegraphics[width=11.5cm]{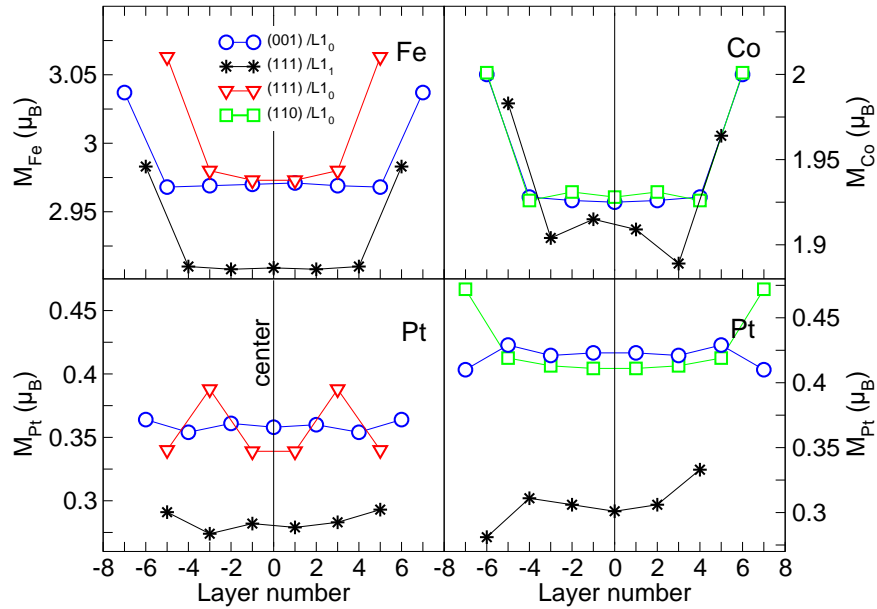}
\caption{(color online) Distribution of spin moments in relaxed FePt and CoPt L1$_0$ and L1$_1$ slabs for various surface orientations. The (111) surface in the L1$_1$ phase is shown as stars (black), all other symbols refer to the L1$_0$ phase: Circles (blue) mark the (001) surface, triangles (red) the (111) surface, and squares (green) belong to the (110) facet in the L1$_0$ phase. For the (111) facet in the L1$_0$ order (mixed atomic composition), we used a seven layer slab which is plotted with doubled layer spacing (always one empty layer in between) for a more convenient comparison with the other surface orientations. Note the different scaling for the Co spin moments, M$_{\rm Co}$, on the right.}
\label{GFePtCoPtLayerMagPAPER}
\end{figure*}
 Our results for the surfaces energies in FePt are collected in the surface-energy phase diagram shown in Fig.~\ref{GsurfDiagrFePtallDRESDENpaper}. The data are in perfect agreement with Hong {\em et al.}\cite{Yoo:05} (see Table \ref{TabelleSurfEbinary}) which were also obtained using the simulation package VASP. In addition we consider facets with two different terminations and surfaces compatible with L1$_1$ order. %With the help of surface-energy phase diagrams we are able to determine a variation range for the surface energy of the single material constituents in binary L1$_0$ or L1$_1$ ordered alloys.
Vertical lines in Fig.~\ref{GsurfDiagrFePtallDRESDENpaper} and Fig.~\ref{GSurfDiagrCoPtallPAPER} mark the limiting cases for the difference in Pt chemical potential $\Delta \mu_{\rm Pt}$ given by the formation enthalpies $\Delta H_{\rm FePt}$ in the L1$_1$ and L1$_0$ phase and the zero value, as for negative values ($\Delta \mu < 0$) no alloying would occur. At the right side of the diagram, a Pt rich environment is assumed and thus $\mu_{\rm Pt} = \mu_{Pt(bulk)}$ and $\Delta \mu_{\rm Pt} = 0$: The surface atoms are in equilibrium with the surrounding Pt-metal and the underlying FePt bulk reservoir. On the left border, at $\Delta H_{\rm FePt} \simeq -0.6$ eV, an Fe-rich environment in the L1$_0$ structure is assumed and therefore $\mu_{\rm Pt} - \mu_{\rm Pt(bulk)} = \Delta H_{\rm FePt}$. The vertical line at $\Delta \mu_{\rm Pt}\simeq -0.4$ eV corresponds to $\Delta H_{\rm FePt}$ in the L1$_1$ phase. The two limiting cases for $\Delta \mu_{\rm Pt}$ (inserted into Eq. \ref{Pt surfEformulaFePtend}) yield the corresponding limiting values for the realistic value of $\sigma_{\rm Pt}$. 
When we artificially vary the stoichiometry by varying $\Delta \mu_{\rm Pt}$ on the horizontal axis, at the same time the difference in Fe chemical potential changes inversely (upper horizontal axis), as can be seen from equation (\ref{EqNonStoich}).\\
The surface energy of Pt covered, i.e. not mixed, (111) facets in the L1$_1$ structure can take extraordinarily low values (in eV/atom):
\begin{equation}\label{platinL11}
0.19 \leq \sigma_{\rm Pt}\leq 0.4.
\end{equation}
These have direct consequences for FePt nanoparticle morphologies and confirm the results of previous {\em ab initio} cluster simulations.\cite{GrunerPartikel:08} These predict radially onion-shell type ordered core-shell icosahedra to be energetically favored over single crystalline L1$_0$ cuboctahedra for small particle diameters. The low energy of Pt-terminated (111) facets overcompensates the high energy of the 20 twins, which possesses an individual L1$_1$ order and the additional contribution of the twin boundaries.\par
For comparison the particle surface energies of orientations with mixed atomic composition as the (100) and the (111) facet in the L1$_0$ phase are given in Fig.~\ref{GsurfDiagrFePtallDRESDENpaper}, too (horizontal lines).\\ 
%These show that the L1$_0$-ordered cuboctahedron, required for data storage applications, is most likely not the thermodynamical stable morphology for cluster sizes below $4 {\rm nm}$.\\
A further promising candidate for future ultra-high density magnetic storage devices is L1$_0$ CoPt due to its similar high magnetocristalline anisotropy energy in the bulk phase (K$_{\rm u} = 5 \cdot10^{7} {\rm erg/cm^{3}}$).
Therefore, analogous calculations were done for CoPt surfaces (cf.~Fig.~\ref{GSurfDiagrCoPtallPAPER}). Qualitatively the same trends for the surface energies were found. Pt-covered (111)-facets in the L1$_1$ structure are here even more favorable (see Table \ref{TabelleSurfEbinary}). 
In general, we find that elemental, solely Pt-terminated surfaces are preferred over Fe covered and mixed surfaces of the same orientation. The consistently low surface energy of Pt covered facets may be regarded as one important driving force for the strong surface segregation tendency of Pt in these alloys.\cite{Ptseg:08,PtsegCuPt:07} 
The lowest energy orientation for mixed surfaces is the highly coordinated (111) surface in accordance with the elemental systems discussed in Section \ref{ElementalSystems}.\\
For binary systems with perfect L1$_0$-order, (111) surfaces which are covered by only one atomic species can not exist for geometric reasons. On the other hand, this surface modification can be realized for the L1$_1$ structure, which is however not stable for bulk FePt. Thus, a sufficiently low surface energy may stabilize the L1$_1$ structure in small particles.
\par
In addition, we have investigated the distribution of the Fe (Co respectively) spin moments and induced Pt moments inside relaxed L1$_0$ and L1$_1$ FePt and CoPt slabs with various surface orientations and different surface terminations (cf.~Fig.~\ref{GFePtCoPtLayerMagPAPER}).
For the case of FePt, we consider an Fe termination for the (001) slab in L1$_0$ order and the (111) slab in L1$_1$ order. (The (111) surface in the L1$_0$ order always consists of mixed atomic composition). For comparison we have chosen for CoPt a Pt-covered (001) and (110) slab in the L1$_0$ phase while the (111) slab in the L1$_1$ phase has one Co and one Pt surface. The (111) and (110) surfaces in L1$_0$ phase are qualitatively the same for both alloys and are thus shown only once.
Again, the spin moment of the transition metal atom is enhanced by about $3-4\%$ at the outermost layers.
% (compare Fig. \ref{GFePtCoPtLayerMagPAPER}, (110), diamonds).\\
 The induced Pt surface moments show a strong dependence of the number of neighboring transition metal atoms. In the outermost layer, the Pt atoms lose a part of their magnetic partners. For the (111) FePt surface in the L1$_0$ phase, which consists of Fe and Pt atoms (mixed atomic composition), the missing magnetic Fe neighbors are decisive for the reduced Pt moment in the outermost layer. In contrast to the situation in the subsurface layer: Here, the Pt atoms have full coordination and thus show an enhanced moment due to the large Fe moment in the surface layer. In a similar manner the Pt moment is slightly enhanced in the subsurface layer of the Fe terminated (001) slab in L1$_0$ order and the (111) slab in L1$_1$ order.
For CoPt this simple rule does not seem to hold true as we find an enhanced Pt moment for the Pt surfaces in the (110) slab in L1$_0$ order as well.
%For the surface orientations with alternating Fe (Co) and Pt layers, i.e. for the (001) and (110) facet in the L1$_0$ phase (blue, rightward triangles and green, upward triangles, respectively) and the (111) facet in the L1$_1$ order (black stars), the Pt moment is reduced in case it resists in the surface layer (again due to lower coordination number) and is enhanced in case it resists in the subsurface layer, in which it has full coordination and feels the enhanced Fe surface moment. 
%\begin{table}
%\begin{ruledtabular}
%\begin{tabular}{c c c c}
%System            &CoPt   & MnPt  & FePt \\ \hline
%$\sigma$(001) &0.977   & 0.986 & 0.991 \\ 
%$\sigma$(111) &0.654   & 0.628 & 0.701 \\
%\end{tabular}
%\end{ruledtabular}
%\caption{Comparison of the averaged surface energies, in eV/atom of the (001) and the (111) facet for the L1$_0$ phase in FePt, MnPt, and CoPt.}
%\label{Compare}
%\end{table}

Recently performed {\em ab initio} cluster simulations revealed that the stability of single crystalline morphologies might be stabilized in these alloys by reducing the number of 3d electrons.\cite{GrunerLatest:08,GrunerPartikel:08,GrunerSub:09} In an extreme case, this may be achieved by changing from FePt to MnPt, as Mn has one 3$d$ electron less than Fe. On the other hand, one has to take care of the strong antiferromagnetic tendencies present in Mn alloys. 
Therefore, we considered the most relevant surfaces (001) and (111) in the L1$_0$ structure of AF MnPt. A collection of calculated surface energies and surface spin moments for FePt, CoPt and MnPt are listed in Table \ref{TabelleSurfEbinary}.
 The surface energies of the (001) facets only vary in between 15 meV/atom for the different binary alloys. 
For the (111)-facet the variation amounts to 70 meV/atom.
Also for MnPt the (111) surface is more favorable than the (001)-facet with an even increasing energy difference. Further investigation of non-stoichiometric ternary Fe-Mn-Pt alloys might thus be an interesting task.\\
 
\subsubsection{Stability range of L1$_0$- versus L1$_1$-ordered clusters}\label{StabRange}
\begin{figure}
\includegraphics[width=7.4cm, angle=-90]{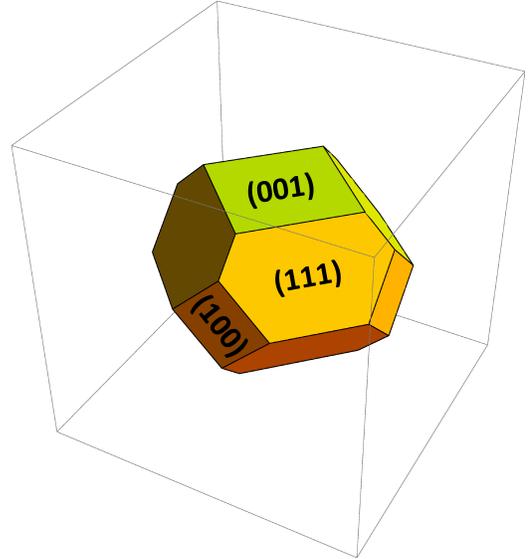}
\caption{(color online) Final shape of the L1$_0$ ordered, asymmetric Wulff polyhedron for the case of FePt with two Pt covered square (001) surfaces (green) on top and bottom, four square (100) facets with mixed atomic composition, and six hexagonal (111) facets at the sides. The distances of the three different facets to the particle center ($\rm d_{001}$, $\rm d_{100}$, and $\rm d_{111}$) are determined following the Wulff construction making use of the calculated surface energies (see Table \ref{TabelleSurfEbinary}). The area of the Pt covered (001) surfaces is considerably enlarged compared to the mixed (100) and (010) surfaces. For CoPt the particle shape is qualitatively the same.}
\label{FePtL10WP}
\end{figure}

\begin{figure}
\includegraphics[width=7.4cm, angle=-90]{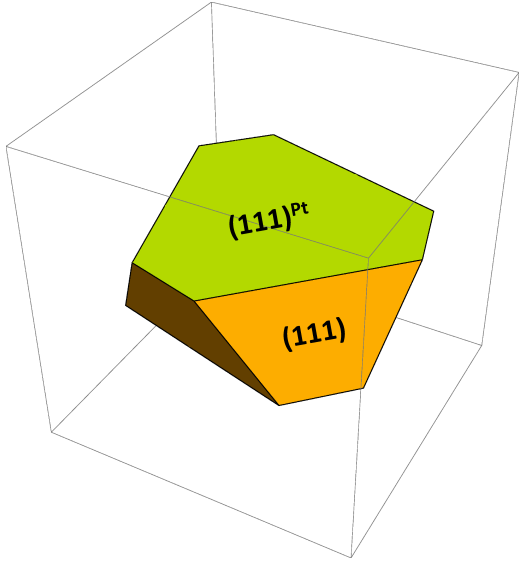}
\caption{(color online) Calculated shape of the L1$_1$ ordered, asymmetric FePt octahedron with two Pt covered, hexagonal (111) Pt surfaces (green) on top and on the bottom, and six hexagonal (111) facets at the sides with mixed atomic composition (orange). The distances of the two different (111) facets to the particle center ($\rm d^{Pt}_{111}$ and $\rm d_{111}$) are determined using the calculated surface energies (see Table \ref{TabelleSurfEbinary}) and applying the Wulff theorem. As in the Wulff polyhedron, the area of the Pt covered (111) surfaces is considerably enlarged due to their extraordinarily low surface energy. For the case of CoPt a similar shape is obtained.}
\label{FePtL11Octa}
\end{figure}

In order to get an idea of the influence of the calculated surface energies on the equilibrium shape of small nanoparticles, we apply a simple approach to approximate the stability range of different structural morphologies in the L1$_0$ and competing L1$_1$ order.
In the limit of large diameters the particles can be regarded as spherical.
For a first rough estimate we make only use of the energy differences between the L1$_0$ and L1$_1$ order for volume and for surface atoms of the lowest energy surfaces. These are the platinum covered (111) facets in the L1$_1$ structure and the platinum covered (001) facets in the L1$_0$ phase. A more realistic picture should also take into account twin boundary energies and internal stress. As the L1$_1$ phase  is not stable for bulk FePt, completely L1$_0$-ordered particles are expected for large diameters. But with decreasing volume, the surface to volume ratio increases and the extraordinarily low surface energy of the platinum covered (111) facet in the L1$_1$ phase gains increasing importance.
Because of this, the L1$_1$-order becomes more favorable than the L1$_0$-ordering for particle sizes below a critical diameter. At this critical diameter, the gain in surface energy is equal to the energy loss due to L1$_1$ ordering. For FePt we find a critical diameter of 3.7\,nm and 6\,nm for CoPt for averaged values of $\sigma$ in agreement with total energy calculations of binary transition metal clusters ( Appendix~\ref{SimpModel} for more details).\cite{GrunerSub:09} \\
 However, the equilibrium crystal shape of an arbitrary particle is not necessarily spherical. 
Following the investigation of the structural stability of single crystalline and multiply twinned FePt nanoparticles, which has recently been performed by M\"uller and Albe,\cite{Albe:07} we apply a more detailed continuum model, in which the different surface energies of the various facets are taken into account in terms of a Wulff construction. M\"uller and Albe considered symmetric particles and surfaces with mixed atomic composition only. We will also allow for asymmetric particle morphologies in the following, details are given in Appendix \ref{detail cont model}.
\begin{figure}
\includegraphics[width=6cm, angle=-90]{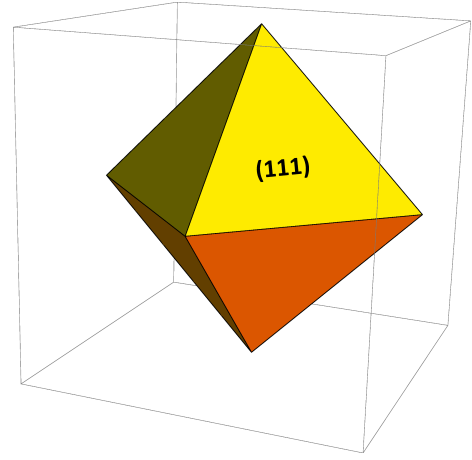}
\caption{(color online) Schematic view of a regular octahedron, terminated solely by eight (111) facets.}
\label{Octahedron}
\end{figure}
Thus, an octahedron which is solely terminated by (111) facets may be considered as a favorable particle morphology (cf.~Fig.~\ref{Octahedron}). Other candidates are multiply twinned morphologies as icosahedra. Here, the calculation of optimum shapes requires the calculation of twinning energies which is beyond the scope of this paper. 
Therefore, we compare only two competing single crystalline structural motifs:
On the one hand, the above mentioned L1$_1$ ordered octahedron with two elemental Pt covered (111) surfaces and six (111) facets of mixed atomic composition (cf.~Fig.~\ref{Octahedron} and Fig.~\ref{FePtL11Octa}) and on the other hand, the L1$_0$ ordered Wulff polyhedron with eight (111) facets of mixed atomic composition, two elemental Pt terminated (001) facets, and four (100) surfaces covered with Fe as well as Pt atoms (cf.~Figure~\ref{FePtL10WP}). 

\begin{figure}
\includegraphics[width=8.4cm]{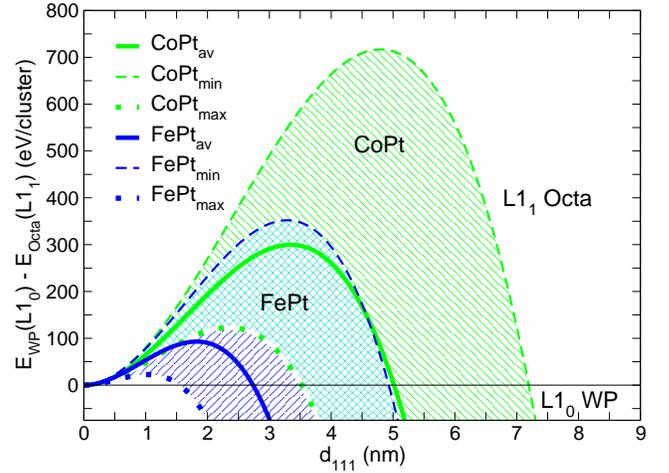}
\caption{(color online) Difference in energy between the L1$_0$ ordered, asymmetric Wulff polyhedron (shown in Fig.~\ref{FePtL10WP}) and the L1$_1$ ordered, asymmetric octahedron (shown in Fig.~\ref{FePtL11Octa}) as a function of the distance $d_{111}$ for CoPt (green) and FePt (blue). The solid lines correspond to the assumption of an averaged surface energy for the Pt terminated (001) facets in L1$_0$ order and Pt terminated (111) facets in L1$_1$ order, the dashed lines belong to maximum and the dotted lines to minimum Pt surface energy values. In the positive regions between the dotted and the dashed lines, i.e. the dark (blue) hatched area for FePt and bright (green) hatched area for CoPt, the L1$_1$ octahedron is the energetically favorable morphology, for negative energy differences, the L1$_1$ Wulff polyhedron is more stable. In the central, double hatched (turquoise) region the FePt and the CoPt regions overlap. Taking averaged Pt surface energy values (solid lines), we find that for distances up to $\rm d_{111} \simeq 2.76$~nm for FePt and $\rm d_{111} \simeq 5$~nm for CoPt (this corresponds to effective particle diameters of $6.28$~nm and $11.26$~nm assuming spherical particles with the respective atomic volume), the L1$_1$ ordered octahedron is lower in energy. Here, the energetic advantage of the L1$_1$ octahedron becomes maximum for $\rm d_{111} \simeq 1.80$~nm for FePt and $\rm d_{111} \simeq 3.35$~nm for CoPt. At this sizes, the L1$_1$ octahedron is about 93 eV/cluster (35.7 meV/atom) for FePt and about 300 eV/cluster (18.2 meV/atom) for CoPt lower in energy than the L1$_0$ Wulff polyhedron.}
\label{GdEL10L11ALL}
\end{figure}

The resulting energy differences, $\rm E^{L1_0}_{WP} -  E^{L1_1}_{Octa}$, are shown in Figure \ref{GdEL10L11ALL} for FePt (blue) and CoPt (green). These are given as a function of $d_{111}$, describing the distance of a (111) facets from the particle center. In addition, we consider the possible variation of the surface energy of Pt terminated (001) facets in L1$_0$ and (111) facets in L1$_1$ order with the chemical potential.
\\
This it is important to mention, since these asymmetric binary structures are in general non-stoichiometric and the composition differs between the two morphologies. Therefore, in a strict sense, we can only give an estimate of the stability range of the different morphologies in the two competing ordered phases. 

For FePt as well as CoPt the asymmetric L1$_1$ ordered octahedron is the energetically preferred particle morphology for sufficiently small particle sizes. Using averaged surface energies of Pt terminated (001) facets in L1$_0$ order and (111) facets in L1$_1$ order, the L1$_1$ octahedron is lower in energy for distances $\rm d_{111} \simeq 2.76$~nm for FePt (blue solid line). This corresponds to particle diameters up to $6.28$~nm (assuming a spherical particle with equal volume) and a total number of approximately $9200$ atoms per cluster. The energetic advantage of the L1$_1$ octahedron becomes maximum for $\rm d_{111} \simeq 1.8$~nm which is equal to a particle diameter of approximately $\simeq 4.12$~nm (about $2600$ atoms) and amounts to $93$~eV/cluster (35.7 meV/atom) (see Figure \ref{GdEL10L11ALL}). 
Applying the same considerations to CoPt leads to a critical particle diameter of approximately $\rm 11.26$~nm below which the L1$_1$ order is the thermodynamically stable phase. At a particle diameter of $\simeq 7.5$~nm, the maximum energy difference ($18.18$ meV/atom) is reached. Here, the CoPt L1$_1$ octahedron contains about $16500$ atoms.
If one takes the minimum possible surface energies for Pt terminated (001) facets in L1$_0$ order and (111) facets in L1$_1$ order, the L1$_1$ octahedron is stable up to $\rm d_{111} \simeq 4.9$~nm for FePt, which corresponds to a critical diameter of $11.18$~nm, and $\rm d_{111} \simeq 7.3$~nm for CoPt (critical diameter of $16.47$~nm). Taking the opposite case, i.e. maximal Pt surface energies, leads to the dotted lines. Here, the L1$_0$ ordered Wulff polyhedron is the thermodynamically stable morphology for particle diameters larger than $\rm d_{111} \simeq1.5$~nm (critical diameter $3.43$~nm) for FePt and $\rm d_{111} \simeq 3.5$~nm (critical diameter $7.9$~nm) for CoPt. In summary, the L1$_1$ ordered, Pt terminated, asymmetric octahedron can be expected for particle diameters between $3.43$~nm and $11.18$~nm for FePt and between $7.9$~nm and $16.47$~nm for CoPt. This large expectation ranges of $\simeq$~7.5\,nm for FePt and $\simeq$~8.5\,nm for CoPt must be seen as a kind of error bar, due to the missing knowledge of the chemical potential of the single material components.\\ 
Nevertheless, in comparison to the simplified considerations above these results yield on average even larger diameters and thus support the prediction that L1$_1$ ordered nanoparticles with Pt covered (111) facets are a competitive particle morphology for small cluster sizes for both, FePt and CoPt binary alloys.
Multiply twinned morphologies as icosahedra, which have been investigated in Ref.~\onlinecite{GrunerPartikel:08}, might even further optimize the area of favorable Pt-covered (111) surfaces, while providing a more spherical shape. This however is achieved at the expense of internal interfaces, which need to be considered separately.
%With respect to the work of K. Albe {\em et al.}\cite{Albe:07}, multiply twinned particles, as i.e. the icosahedron, are expected to be the most favorable particle morphology for very small particle sizes ($\leq 2.7\, {\rm nm}$ in diameter). This is, among others, caused by the fact that the icosahedra has a more spherical shape than the Wulff polyeder and the octahedra. This surface shape optimization however,
% is done at the expense of internal stress and twin boundary energies and thus only occurs for sufficiently small particle sizes.\\
%But one question remains open: The Wulff construction is valid for a macroscopic crystal, but for particle diameters around $4\, \rm nm$ the applicability of a continuum model might be questioned. The values obtained for the surface energy of macroscopic surfaces might no longer be valid and furthermore, the edges separating the different facets are composed of atoms with lower coordination (1D) than the atoms in the facets themselves. Therefore these atomic surface energies are higher and should be taken into account. For very small particle sizes a refined (atomistic) model might be required.  
\par

\section{Conclusion}
We have calculated surface energies and surface magnetism of various low indexed surfaces for the elemental systems bcc and fcc Fe, fcc Co, fcc Pt and for the binary (fct) alloys FePt, CoPt, and MnPt. For Fe, Co and Pt we have considered the (001), (110) and the (111) surfaces. In addition, also the (100), (011) orientation in the L1$_0$ phase and the (111) surface in the L1$_1$ structure have been examined for the binary alloys.
The surface energies were determined using the slab approach. For the special surface orientations in the binary alloys with more than one possible coverage, surface-energy phase diagrams have been evaluated in order to account for the surface energy contributions of the single material components.\\
For all systems under investigation, (111) facets show the lowest surface energy. Especially Pt covered (111) surfaces, as found in L1$_1$ ordered FePt, possess an extraordinarily low surface energy which is considerably lower than the respective surface energy of pure Pt. This gives rise to the preferred appearance of Pt terminated core-shell icosahedral nanoparticles in gas-phase experiments and agrees well with the results of theoretical cluster calculations which show that platinum terminated, radially L1$_0$-ordered core-shell icosahedra are energetically favorable.\cite{GrunerPartikel:08,GrunerSub:09}\\
The surface energies of CoPt qualitatively follow the same trend as found for FePt: 
 $\gamma\,(L1_1/111)\,\leq\,\gamma\,(111)\,<\,\gamma\,(100)\,\leq\,\gamma\,(001)\,<\,\gamma\,(011)\,\leq\,\gamma\,(110)$. For L1$_0$ MnPt, the surface energies of the (001) and the (111) facet lie in the same range as those of FePt and CoPt while (111) surfaces are still previleged. We may speculate that the addition of Mn to FePt should not substantially modify the relation of the surface energies. The surface energy of purely Pt-covered surfaces is always lower than the energy of the elemental surfaces of the corresponding 3d-metal.\\
These results allow us to estimate the stability range of the most favorable particle morphologies in L1$_0$ and L1$_1$ order. In a first simple approach, relying only on the energy difference between the L1$_1$ and the L1$_0$ structure for bulk and for surface atoms, we can derive a critical diameter below which the L1$_1$ phase may be stabilized. For FePt, this diameter is about $3.7\, \rm nm$, which is in good agreement with {\em ab initio} cluster simulations.\cite{GrunerSub:09} For CoPt we find a critical diameter of $6\, \rm nm$.
 Similar crossover sizes were obtained within a refined continuum model, which allows to assess contributions of different faces more precisely. Assuming single crystalline particles, candidates for stable structures can be determined following the Wulff construction: Favorable single crystalline morphologies are the non-spherical Wulff polyhedron with Pt covered (001) facets for the L1$_0$ order and the asymmetric octahedron with Pt covered (111) facets in case of the L1$_1$ order. Comparing the total energy of those two structural motifs and using averaged Pt chemical potential yields a critical diameter of $\simeq 6$ nm diameter for FePt and $\simeq 11$ nm diameter for CoPt below which the L1$_1$ ordered, asymmetric octahedron is the energetically preferred structure.\\ 
 %This result agrees well with static monte carlo calculations of K. Albe {\em et al.}
This underlines the central result of this study, that the extraordinarily low surface energy of elemental Pt terminated (111) facets of L1$_1$ bulk crystals may stabilize FePt and CoPt nanoparticles with L1$_1$ crystalline order for sufficiently small particle diameters - although the corresponding bulk materials are unstable in the L1$_1$ structure.\\

\acknowledgments
The authors would like to thank P. Kratzer and H. C. Herper for fruitful discussions.
Financial support was granted by the Deutsche Forschungsgemeinschaft (SFB 445 and SPP 1239).

\begin{appendix}\label{appendix} 
\section{Spherical model for the clusters}\label{SimpModel}
In the limit of large diameters the particles are assumed to be spherical and the difference in surface energy per atom is approximated by $\Delta E_{\rm S} = \sigma_{\rm Pt}^{L1_0}(111) - \sigma_{\rm Pt}^{L1_1}(001)$. The volume energy difference per atom between the L1$_1$ phase and the L1$_0$ phase is $\Delta E_{\rm V} = E_{\rm bulk}^{L1_0} - E_{\rm bulk}^{L1_1}$. The gain in surface energy is equal to the energy loss due to L1$_1$ ordering if the following condition is fullfilled:
\begin{equation}\label{condition}
\Delta E= N \cdot \Delta E_{\rm V}+  S\cdot \Delta E_{\rm S} = 0.
\end{equation}
Here $N $ is the total number of atoms in the particle and $S$ the number of surface atoms. Relation~(\ref{condition}) gives the percentage of surface atoms that leads to the stability of the L1$_1$ phase:
\begin{equation}\label{percentage}
-\frac{\Delta E_{\rm V}}{\Delta E_{\rm S}} = \frac{S}{N } = \delta_{\rm S}.
\end{equation}
With the bulk and surface energy differences taken from section \ref{SurfFePt}, $\Delta E_{\rm V} = -0.13\, {\rm eV/atom}$ and $\Delta E_{\rm S} = 0.39\,  {\rm eV/atom}$ for FePt, we estimate that the critical percentage of surface atoms per particle ($\delta_{\rm S}$) amounts to 32.7\%. We can find the corresponding total number of atoms per particle $N $ using $\delta_{\rm S}=4/\sqrt[3]{N }$.\cite{Edelstein:97} For $\delta_{\rm S} = 0.327$ this leads to $N $$\simeq$\,1840. The particle diameter can then be estimated with the help of the averaged atomic volume in L1$_0$ FePt, $\Omega$, using the formula $N \Omega = V_{\rm sphere} = (4\pi/3)r^3$.

\section{Asymmetric continuum model}\label{detail cont model}\label{L10}
As the surface energies of the various facets in L1$_0$- and L1$_1$-ordered FePt and CoPt differ considerably, the assumption of a spherical particle is not necessarily valid. Rather asymmetric particles with an enlarged area ot the energetically favored facets are expected. This is taken care of in a more detailed continuum model.
 For single crystalline metal particles, the thermodynamically stable shape is determined by a Wulff construction\cite{Wulff:01}, where the energy minimizing shape is given by a constant ratio $\gamma_{\rm hkl}/d_{\rm hkl}$ with $d_{\rm hkl}$ the distance from particle center of a (hkl) facet with surface energy per unit area $\gamma_{\rm hkl}$. The Wulff theorem applies to a macroscopic crystal.
The ratios $\Gamma_{100} = \gamma_{100} / \gamma_{111}$ and $\Gamma_{001} = \gamma^{\rm Pt}_{001} / \gamma_{111}$ for the L1$_0$ structure, and $\tilde{\Gamma} = \gamma^{\rm Pt}_{111} / \gamma^{\rm mix}_{111}$ for the L1$_1$ phase determine which particle morphology possesses the energy minimizing shape. In the strong faceting limit ($\gamma_{\rm 110} / \gamma_{\rm 111} \geq \sqrt{3/2}$) and if the condition $\sqrt{3}/2 \leq \Gamma_{100} \leq \sqrt{3}$ is fulfilled, the Wulff shape is a truncated octahedron terminated by (111) and (100) facets only (cf.~Fig.~\ref{FePtL10WP}). This applies to L1$_0$ FePt and CoPt particles where $\Gamma_{100} = 1.205$ and $\Gamma_{100} = 1.265$, respectively.
In the limiting case of $\Gamma_{100} = \sqrt{3} \simeq 1.732$ the Wulff construction leads to a regular octahedron as depicted in Fig. \ref{Octahedron}. 
The L1$_0$ ordered Wulff polyhedron can be constructed with the help of the ratios $\Gamma_{\rm 100} = \gamma_{\rm 100} / \gamma_{\rm 111}$ and $\Gamma_{\rm 001} = \gamma^{\rm Pt}_{\rm 001} / \gamma_{\rm 111}$  by truncating the vertices of a regular octahedron at distances $\rm d_{111}$, $\rm d_{100}$ and $\rm d_{001}$ from the center.\cite{Wulff:01,Albe:07} The volume $V$ of the truncated octahedron can easily be derived by subtracting from the total octahedron volume, $V_{\rm Octa}$, half of the volume of two small octahedra truncated at the vertices in [001] direction, $V^{\rm d_{001}}_{\rm Octa}$, and four small octahedra truncated in [100] directions, $V^{\rm d_{100}}_{\rm Octa}$:
\begin{equation}\label{Veasy}
\rm V = V_{\rm Octa}-\left(V^{d_{001}}_{\rm Octa} + 2 V^{d_{100}}_{\rm Octa} \right)
\end{equation}
with $V_{\rm Octa} = 4 \sqrt{3} \,\rm d^3_{111}$ and $V^{\rm d_{100}}_{\rm Octa} = 4 \sqrt{3} \,(\rm d_{111}-\frac{\rm d_{100}}{\sqrt{3}})^3 $ (analogously for $V^{\rm d_{001}}_{\rm Octa}$).
The area of the (100) and (001) surfaces is simply given by the square of the edge length while the remaining (111) surface areas of the octahedron can be calculated following the same idea when determining the volume:
\begin{equation}\label{Aeasy}
\rm A_{111} = \frac{1}{8} \left( A_{\rm Octa}- (A^{d_{001}}_{\rm Octa} + 2A^{d_{100}}_{\rm Octa}) \right)
\end{equation}
with
\begin{equation} 
A_{\rm Octa}= 12 \sqrt{3} \rm d^2_{111}
\end{equation}
and 
\begin{equation}
A^{\rm d_{001}}_{\rm Octa} = 12 \sqrt{3} (\rm d_{111} - \frac{d_{001}}{\sqrt{3}})^2,
\end{equation}
(correspondingly for $A^{\rm d_{001}}_{\rm Octa} $).
 \\
With this preliminary considerations and after substituting $\rm d_{100} = \Gamma_{100} d_{111}$ and $\rm d_{001} = \Gamma_{001} d_{111}$, the volume $V$ and the total area $A_{\rm hkl}$ of (111), (100) and Pt covered (001) surfaces of the L1$_0$ Wulff polyhedron can be written as a function of the variable $\rm d_{111}$:
\begin{equation}\label{Volume1}
\rm V(d_{111}) = 4 d_{111}^3 \sqrt{3} \left[1-\left(1-\frac{\Gamma_{001}}{\sqrt{3}} \right)^3 - 2 \left(1-\frac{\Gamma_{100}}{\sqrt{3}} \right)^3 \right],
\end{equation}
\begin{equation}\label{Area001}
\rm A_{001} (d_{111})=  6 d_{111}^2 \left(1- \frac{\Gamma_{001}}{\sqrt{3}} \right)^2,
\end{equation}
\begin{equation}\label{Area100}
\rm A_{100} (d_{111})=  6 d_{111}^2 \left(1- \frac{\Gamma_{100}}{\sqrt{3}} \right)^2,
\end{equation}
%$$
%\begin{array}{rr}\label{Area111}
\begin{equation}\label{Area111}
\rm A_{111} (d_{111})=  \frac{3}{2} \sqrt{3} d_{111}^2 \left[ 1- \left(1-\frac{\Gamma_{001}}{\sqrt{3}} \right)^2 - 2 \left(1-\frac{\Gamma_{100}}{\sqrt{3}} \right)^2 \right].
\end{equation}
As the surface energy of Pt terminated (001) facets, $\gamma^{\rm Pt}_{001}$, is a function of the difference in Pt chemical potentials, $\rm \mu_{Pt}~-~\mu_{Pt(bulk)}$, (Fig.~\ref{GsurfDiagrFePtallDRESDENpaper} and Fig.~\ref{GSurfDiagrCoPtallPAPER}) the value of $\Gamma_{\rm 001} ~=~ \gamma^{\rm Pt}_{\rm 001} / \gamma_{\rm 111}$ also varies  between two limiting cases:
\begin{equation}\label{G001Vary}
\rm \Gamma^{min}_{\rm 001} = \frac{\gamma^{\rm Pt,min}_{\rm 001}}{\gamma_{\rm 111}} \leq \Gamma_{\rm 001} \leq \Gamma^{max}_{\rm 001} = \frac{\gamma^{\rm Pt,max}_{\rm 001}}{\gamma_{\rm 111}}
\end{equation}
leading to $0.686 \leq \Gamma_{\rm 001} \leq 1.074$ for FePt and $0.816 \leq \Gamma_{\rm 001} \leq 0.950$ for CoPt.
 The averaged values are almost the same for FePt and CoPt: $\rm \Gamma^{\rm av}_{\rm 001} = 0.88$ and $\rm \Gamma^{\rm av}_{\rm 001} = 0.883$, respectively. Thus, also the volume, $V$, and the areas $A_{001}$ and $A_{111}$ are not exactly determined but vary as a function of $\rm \mu_{Pt}-\mu_{Pt(bulk)}$. For the quantitative examples and representative particle shapes shown in section \ref{StabRange} averaged values for $\rm \Gamma_{\rm 001}$ are used.\\
 
 \label{L11}
The non-spherical, asymmetric L1$_1$-ordered octahedron is defined by the ratios
\begin{equation}\label{GamL11}
\rm \tilde{\Gamma} = \frac{\gamma^{Pt}_{111}}{\gamma^{mix}_{111}} = 0.444\,\,for\,\,FePt
\end{equation}
and
\begin{equation}\label{GamL11CoPt}
\rm \tilde{\Gamma}  = \frac{\gamma^{Pt}_{111}}{\gamma^{mix}_{111}}  = 0.446\,\, for\,\,CoPt.
\end{equation}
As for Pt terminated (001) facets in L1$_0$ order, here, the Pt covered (111) facets are known only in between two limiting cases (Fig.~\ref{GsurfDiagrFePtallDRESDENpaper} and Fig.~\ref{GSurfDiagrCoPtallPAPER}) yielding: $0.277 ~\leq~ \rm \tilde{\Gamma} ~\leq~ 0.590$ for FePt and $0.338 ~\leq~ \rm \tilde{\Gamma} ~\leq~ 0.503$ for CoPt.
In analogy to the L1$_0$ ordered Wulff polyhedron, we express the volume $\rm \tilde{V}$, the total area of Pt covered (111) facets $\rm \tilde{A}^{Pt}_{111}$, and (111) facets of mixed composition $\rm \tilde{A}^{mix}_{111}$ for the L1$_1$ ordered octahedron as a function of $\rm d_{111}$ and $\rm \tilde{\Gamma}$ by
\begin{equation}\label{Volume2}
\rm \tilde{V}(d_{111}) = \frac{1}{2} \sqrt{3} \, d_{111}^3 \tilde{\Gamma} \left(9 - \tilde{\Gamma}^2 \right),
\end{equation}
\begin{equation}\label{Axy}
\rm \tilde{A}^{mix}_{111} (d_{111})=  \frac{3}{2} \sqrt{3} \, d_{111}^2 \tilde{\Gamma},
\end{equation}
\begin{equation}\label{Az}
\rm \tilde{A}^{Pt}_{111} (d_{111})=  \frac{3}{4} \sqrt{3} \, d_{111}^2 \left( 3 -  \tilde{\Gamma}^2 \right).
\end{equation}
Under the assumption of an averaged value of the surface energy, $\rm \gamma^{Pt,av}_{111}=0.733$ for FePt and $\rm \gamma^{Pt,av}_{111}=0.722$ for CoPt we were able to precise the structural motifs. The resulting particle shapes are shown in Fig.~\ref{FePtL10WP} and \ref{FePtL11Octa}, respectively.
\\
As expected, the Pt covered (001) facets are considerably enlarged by almost a factor of $\rm A_{001}/A_{100}=2.6$ compared to the (100) facets with mixed atomic composition ($\rm A_{001}/A_{100}=3.3$ for CoPt). The ratio of the distance of the Pt terminated (001) facet and the mixed (100) facet from particle center ($\rm d_{001}$ and $\rm d_{100}$, respectively) giving the aspect ratio of the particle, amounts to $\rm d_{001}/d_{100}=0.73$ ($\rm d_{001}/d_{100} = 0.70$ for CoPt).\\
Using the continuum model, also the total energy of a particle can be expressed as a function of the distance of a (111) facet from the particle center $d_{111}$. For large enough particles, it can be approximated by the sum of volume and surface energy terms.\cite{Albe:07} 
If $N $ denotes the total number of atoms in a particle, the particle volume is given by $V = N  \Omega$, where $\Omega$ is the atomic volume. Thus, the total number of atoms varies with cluster size as
\begin{equation}\label{N}
N (d_{111}) = V(d_{111})/ \Omega.
\end{equation}
If we neglect the twin boundary energy and the contributions of edge and corner atoms, we obtain: 
\begin{equation}\label{EContModel}
E(d_{111}) = N (d_{111}) E_{\rm bulk} + \sum_{\rm hkl} A_{hkl}(d_{111}) \gamma_{\rm hkl}.
\end{equation}
Applying Eq. (\ref{EContModel}) yields for the L1$_0$ Wulff polyhedron (WP)
\begin{equation}\label{EWulffPoly}
E_{\rm WP}^{\rm L1_0} = N E_{\rm bulk}^{\rm L1_0} + 8 A_{111} \gamma_{111} + 2 A_{001} \gamma^{\rm Pt}_{001} + 4 A_{100} \gamma^{\rm mix}_{100},
\end{equation}
The dependence on $d_{111}$ is formally obmitted for simplicity. As mentioned above, also the total energy is a function of the difference in Pt chemical potentials, $\rm \mu_{Pt}~-~\mu_{Pt(bulk)}$, and is determined only in a certain range. 
\\
Analogously such considerations also apply for the L1$_1$ ordered octahedron. Again, the low surface energy of Pt covered (111) facets will lead to an enlargement, compared to (111) facets with mixed atomic composition. We here also consider a non-spherical, asymmetric shape. 
Applying Eq. (\ref{EContModel}) to the L1$_1$ ordered, asymmetric octahedron gives
\begin{equation}\label{EOcta}
E_{\rm Octa}^{\rm L1_1} = N E_{\rm bulk}^{\rm L1_1} + 2 \tilde{A}^{\rm Pt}_{111} \tilde{\gamma}^{\rm Pt}_{111} + 6 \tilde{A}_{111} \tilde{\gamma}_{111}.
\end{equation}
Under the assumption of an averaged value of the surface energy for FePt and CoPt we are able to predict structural motifs. Their shape is shown in Fig.~\ref{FePtL11Octa}. Indeed, the hexagonal Pt terminated (111) facets on top and bottom are enlarged by a factor of $\rm A^{Pt}_{111}/A^{mix}_{111}=3.44$ for FePt ($\rm A^{Pt}_{111}/A^{mix}_{111}= 3.35$ for CoPt) compared to the (111) facets of mixed atomic composition on the side of the particle. The distance of the Pt terminated (001) facet from particle center, $\rm d_{001}$, is even more shortened compared to the mixed (100) facet, $\rm d_{100}$, as found for the L1$_0$ WP. The ratio amounts to $\rm d^{Pt}_{111}/d_{111}=0.412$ for FePt and $\rm d^{Pt}_{111}/d_{111}=0.421$ for CoPt.\\

\end{appendix}

%\input{ErstPapLiteratur.tex}

%\bibliographystyle{apsrev}
%\bibliography{example}

\begin{thebibliography}{0}
\expandafter\ifx\csname natexlab\endcsname\relax\def\natexlab#1{#1}\fi
\expandafter\ifx\csname bibnamefont\endcsname\relax
  \def\bibnamefont#1{#1}\fi
\expandafter\ifx\csname bibfnamefont\endcsname\relax
  \def\bibfnamefont#1{#1}\fi
\expandafter\ifx\csname citenamefont\endcsname\relax
  \def\citenamefont#1{#1}\fi
\expandafter\ifx\csname url\endcsname\relax
  \def\url#1{\texttt{#1}}\fi
\expandafter\ifx\csname urlprefix\endcsname\relax\def\urlprefix{URL }\fi
\providecommand{\bibinfo}[2]{#2}
\providecommand{\eprint}[2][]{\url{#2}}

\end{thebibliography}


\begin{thebibliography}{100}
\expandafter\ifx\csname natexlab\endcsname\relax\def\natexlab#1{#1}\fi
\expandafter\ifx\csname bibnamefont\endcsname\relax
  \def\bibnamefont#1{#1}\fi
\expandafter\ifx\csname bibfnamefont\endcsname\relax
  \def\bibfnamefont#1{#1}\fi
\expandafter\ifx\csname citenamefont\endcsname\relax
  \def\citenamefont#1{#1}\fi
\expandafter\ifx\csname url\endcsname\relax
  \def\url#1{\texttt{#1}}\fi
\expandafter\ifx\csname urlprefix\endcsname\relax\def\urlprefix{URL }\fi
\providecommand{\bibinfo}[2]{#2}
\providecommand{\eprint}[2][]{\url{#2}}

\bibitem[{\citenamefont{Perez et~al.}(2005)\citenamefont{Perez, Dupuis,
  Tuaillon-Combes, Bardotti, Pr\'evel, Bernstein, M\'elinon, Favre, Hannour,
  and Lamet}}]{Perez:05}
\bibinfo{author}{\bibfnamefont{A.}~\bibnamefont{Perez}},
  \bibinfo{author}{\bibfnamefont{V.}~\bibnamefont{Dupuis}},
  \bibinfo{author}{\bibfnamefont{J.}~\bibnamefont{Tuaillon-Combes}},
  \bibinfo{author}{\bibfnamefont{L.}~\bibnamefont{Bardotti}},
  \bibinfo{author}{\bibfnamefont{B.}~\bibnamefont{Pr\'evel}},
  \bibinfo{author}{\bibfnamefont{E.}~\bibnamefont{Bernstein}},
  \bibinfo{author}{\bibfnamefont{P.}~\bibnamefont{M\'elinon}},
  \bibinfo{author}{\bibfnamefont{L.}~\bibnamefont{Favre}},
  \bibinfo{author}{\bibfnamefont{A.}~\bibnamefont{Hannour}}, \bibnamefont{and}
  \bibinfo{author}{\bibfnamefont{M.}~\bibnamefont{Lamet}},
  \bibinfo{journal}{Adv. Eng. Mater.} \textbf{\bibinfo{volume}{7}},
  \bibinfo{pages}{475} (\bibinfo{year}{2005}).

\bibitem[{\citenamefont{Yang et~al.}(2004)\citenamefont{Yang, Liu, Yu, Klemmer,
  Johns, and Weller}}]{Yang:04}
\bibinfo{author}{\bibfnamefont{X.}~\bibnamefont{Yang}},
  \bibinfo{author}{\bibfnamefont{C.}~\bibnamefont{Liu}},
  \bibinfo{author}{\bibfnamefont{J.}~\bibnamefont{Yu}},
  \bibinfo{author}{\bibfnamefont{T.}~\bibnamefont{Klemmer}},
  \bibinfo{author}{\bibfnamefont{E.}~\bibnamefont{Johns}}, \bibnamefont{and}
  \bibinfo{author}{\bibfnamefont{D.}~\bibnamefont{Weller}},
  \bibinfo{journal}{J. Vac. Sci. Technol. B} \textbf{\bibinfo{volume}{22}},
  \bibinfo{pages}{31} (\bibinfo{year}{2004}).

\bibitem[{\citenamefont{Weller and Moser}(1999)}]{Moser:99}
\bibinfo{author}{\bibfnamefont{D.}~\bibnamefont{Weller}} \bibnamefont{and}
  \bibinfo{author}{\bibfnamefont{A.}~\bibnamefont{Moser}},
  \bibinfo{journal}{IEEE Trans. Magn.} \textbf{\bibinfo{volume}{35}},
  \bibinfo{pages}{4423} (\bibinfo{year}{1999}).

\bibitem[{\citenamefont{Lyubina et~al.}(2005)\citenamefont{Lyubina, Opahle,
  M\"uller, Gutfleisch, Richter, Wolf, and Schultz}}]{Opahle:05}
\bibinfo{author}{\bibfnamefont{J.}~\bibnamefont{Lyubina}},
  \bibinfo{author}{\bibfnamefont{I.}~\bibnamefont{Opahle}},
  \bibinfo{author}{\bibfnamefont{K.-H.} \bibnamefont{M\"uller}},
  \bibinfo{author}{\bibfnamefont{O.}~\bibnamefont{Gutfleisch}},
  \bibinfo{author}{\bibfnamefont{M.}~\bibnamefont{Richter}},
  \bibinfo{author}{\bibfnamefont{M.}~\bibnamefont{Wolf}}, \bibnamefont{and}
  \bibinfo{author}{\bibfnamefont{L.}~\bibnamefont{Schultz}},
  \bibinfo{journal}{J. Phys.: Condens. Matter} \textbf{\bibinfo{volume}{17}},
  \bibinfo{pages}{4157} (\bibinfo{year}{2005}).

\bibitem[{\citenamefont{Honolka et~al.}(2009)\citenamefont{Honolka, Lee,
  Kuhnke, Enders, Skomski, Bornemann, Mankovsky, Minar, Staunton, Ebert
  et~al.}}]{Ebert:09}
\bibinfo{author}{\bibfnamefont{J.}~\bibnamefont{Honolka}},
  \bibinfo{author}{\bibfnamefont{T.~Y.} \bibnamefont{Lee}},
  \bibinfo{author}{\bibfnamefont{K.}~\bibnamefont{Kuhnke}},
  \bibinfo{author}{\bibfnamefont{A.}~\bibnamefont{Enders}},
  \bibinfo{author}{\bibfnamefont{R.}~\bibnamefont{Skomski}},
  \bibinfo{author}{\bibfnamefont{S.}~\bibnamefont{Bornemann}},
  \bibinfo{author}{\bibfnamefont{S.}~\bibnamefont{Mankovsky}},
  \bibinfo{author}{\bibfnamefont{J.}~\bibnamefont{Minar}},
  \bibinfo{author}{\bibfnamefont{J.}~\bibnamefont{Staunton}},
  \bibinfo{author}{\bibfnamefont{H.}~\bibnamefont{Ebert}},
  \bibnamefont{et~al.}, \bibinfo{journal}{Phys. Rev. Lett.}
  \textbf{\bibinfo{volume}{102}}, \bibinfo{pages}{067207}
  (\bibinfo{year}{2009}).

\bibitem[{\citenamefont{K\"odderitzsch
  et~al.}(2007)\citenamefont{K\"odderitzsch, Ebert, Rowlands, and
  Ernst}}]{KKRFePt:07}
\bibinfo{author}{\bibfnamefont{D.}~\bibnamefont{K\"odderitzsch}},
  \bibinfo{author}{\bibfnamefont{H.}~\bibnamefont{Ebert}},
  \bibinfo{author}{\bibfnamefont{D.~A.} \bibnamefont{Rowlands}},
  \bibnamefont{and} \bibinfo{author}{\bibfnamefont{A.}~\bibnamefont{Ernst}},
  \bibinfo{journal}{New J. Physics} \textbf{\bibinfo{volume}{9}},
  \bibinfo{pages}{81} (\bibinfo{year}{2007}).

\bibitem[{\citenamefont{Shick and Mryasov}(2003)}]{Shick:03}
\bibinfo{author}{\bibfnamefont{A.~B.} \bibnamefont{Shick}} \bibnamefont{and}
  \bibinfo{author}{\bibfnamefont{O.~N.} \bibnamefont{Mryasov}},
  \bibinfo{journal}{Phys. Rev. B} \textbf{\bibinfo{volume}{67}},
  \bibinfo{pages}{172407} (\bibinfo{year}{2003}).

\bibitem[{\citenamefont{Gruner}(2008)}]{Gruner:08}
\bibinfo{author}{\bibfnamefont{M.~E.} \bibnamefont{Gruner}},
  \bibinfo{journal}{J. Phys. D: Appl. Phys.} \textbf{\bibinfo{volume}{41}},
  \bibinfo{pages}{134015} (\bibinfo{year}{2008}).

\bibitem[{\citenamefont{Zotov and Ludwig}(2008)}]{Zotov:08}
\bibinfo{author}{\bibfnamefont{N.}~\bibnamefont{Zotov}} \bibnamefont{and}
  \bibinfo{author}{\bibfnamefont{A.}~\bibnamefont{Ludwig}},
  \bibinfo{journal}{Intermetallics} \textbf{\bibinfo{volume}{16}},
  \bibinfo{pages}{113} (\bibinfo{year}{2008}).

\bibitem[{\citenamefont{MacLaren et~al.}(2005)\citenamefont{MacLaren,
  Duplessis, Stern, and Willoughby}}]{MacLaren:05}
\bibinfo{author}{\bibfnamefont{J.~M.} \bibnamefont{MacLaren}},
  \bibinfo{author}{\bibfnamefont{R.~R.} \bibnamefont{Duplessis}},
  \bibinfo{author}{\bibfnamefont{R.~A.} \bibnamefont{Stern}}, \bibnamefont{and}
  \bibinfo{author}{\bibfnamefont{S.}~\bibnamefont{Willoughby}},
  \bibinfo{journal}{IEEE Trans. Magn.} \textbf{\bibinfo{volume}{41}},
  \bibinfo{pages}{4374} (\bibinfo{year}{2005}).

\bibitem[{\citenamefont{Podg\'orny}(1991)}]{Podgorny:91}
\bibinfo{author}{\bibfnamefont{M.}~\bibnamefont{Podg\'orny}},
  \bibinfo{journal}{Phys. Rev. B} \textbf{\bibinfo{volume}{43}},
  \bibinfo{pages}{11300} (\bibinfo{year}{1991}).

\bibitem[{\citenamefont{Solovyev et~al.}(1995)\citenamefont{Solovyev,
  Dederichs, and Mertig}}]{Mertig:95}
\bibinfo{author}{\bibfnamefont{I.~V.} \bibnamefont{Solovyev}},
  \bibinfo{author}{\bibfnamefont{P.~H.} \bibnamefont{Dederichs}},
  \bibnamefont{and} \bibinfo{author}{\bibfnamefont{I.}~\bibnamefont{Mertig}},
  \bibinfo{journal}{Phys. Rev. B} \textbf{\bibinfo{volume}{52}},
  \bibinfo{pages}{13419} (\bibinfo{year}{1995}).

\bibitem[{\citenamefont{Antoniak et~al.}(2008)\citenamefont{Antoniak, Trunova,
  Spasova, Farle, Wende, Wilhelm, and Rogalev}}]{Caroline:08}
\bibinfo{author}{\bibfnamefont{C.}~\bibnamefont{Antoniak}},
  \bibinfo{author}{\bibfnamefont{A.}~\bibnamefont{Trunova}},
  \bibinfo{author}{\bibfnamefont{M.}~\bibnamefont{Spasova}},
  \bibinfo{author}{\bibfnamefont{M.}~\bibnamefont{Farle}},
  \bibinfo{author}{\bibfnamefont{H.}~\bibnamefont{Wende}},
  \bibinfo{author}{\bibfnamefont{F.}~\bibnamefont{Wilhelm}}, \bibnamefont{and}
  \bibinfo{author}{\bibfnamefont{A.}~\bibnamefont{Rogalev}},
  \bibinfo{journal}{Phys. Rev. B} \textbf{\bibinfo{volume}{78}},
  \bibinfo{pages}{041406(R)} (\bibinfo{year}{2008}).

\bibitem[{\citenamefont{Friedenberger}(2007)}]{Nina:07}
\bibinfo{author}{\bibfnamefont{N.}~\bibnamefont{Friedenberger}},
  \emph{\bibinfo{title}{{Layer resolved Lattice Relaxation in magnetic $\rm
  Fe_xPt_{1-x}$ Nanopartices}}}, \bibinfo{howpublished}{Universit\"at
  Duisburg-Essen} (\bibinfo{year}{2007}), \bibinfo{note}{{\rm Diploma thesis}}.

\bibitem[{\citenamefont{Dmitrieva}(2007)}]{Olga:07}
\bibinfo{author}{\bibfnamefont{O.}~\bibnamefont{Dmitrieva}}, Ph.D. thesis,
  \bibinfo{school}{Universit\"at Duisburg-Essen} (\bibinfo{year}{2007}).

\bibitem[{\citenamefont{Sudfeld et~al.}(2007)\citenamefont{Sudfeld, Dmitrieva,
  Friedenberger, Dumpich, Farle, Song, Kisielowski, Gruner, and
  Entel}}]{NinaPaper:07}
\bibinfo{author}{\bibfnamefont{D.}~\bibnamefont{Sudfeld}},
  \bibinfo{author}{\bibfnamefont{O.}~\bibnamefont{Dmitrieva}},
  \bibinfo{author}{\bibfnamefont{N.}~\bibnamefont{Friedenberger}},
  \bibinfo{author}{\bibfnamefont{G.}~\bibnamefont{Dumpich}},
  \bibinfo{author}{\bibfnamefont{M.}~\bibnamefont{Farle}},
  \bibinfo{author}{\bibfnamefont{C.~Y.} \bibnamefont{Song}},
  \bibinfo{author}{\bibfnamefont{C.}~\bibnamefont{Kisielowski}},
  \bibinfo{author}{\bibfnamefont{M.~E.} \bibnamefont{Gruner}},
  \bibnamefont{and} \bibinfo{author}{\bibfnamefont{P.}~\bibnamefont{Entel}},
  \bibinfo{journal}{Mater. Res. Soc. Symp. Proc.}
  \textbf{\bibinfo{volume}{998E}}, \bibinfo{pages}{0998}
  (\bibinfo{year}{2007}).

\bibitem[{\citenamefont{Dai et~al.}(2002)\citenamefont{Dai, Sun, and
  Wang}}]{Wang:02}
\bibinfo{author}{\bibfnamefont{Z.~R.} \bibnamefont{Dai}},
  \bibinfo{author}{\bibfnamefont{S.}~\bibnamefont{Sun}}, \bibnamefont{and}
  \bibinfo{author}{\bibfnamefont{Z.~L.} \bibnamefont{Wang}},
  \bibinfo{journal}{Surf. Sci.} \textbf{\bibinfo{volume}{505}},
  \bibinfo{pages}{325} (\bibinfo{year}{2002}).

\bibitem[{\citenamefont{Wang et~al.}(2008)\citenamefont{Wang, Dmitrieva, Farle,
  Dumpich, Farle, Ye, Poppa, Kilaas, and Kisielowski}}]{Wang:08}
\bibinfo{author}{\bibfnamefont{R.~M.} \bibnamefont{Wang}},
  \bibinfo{author}{\bibfnamefont{O.}~\bibnamefont{Dmitrieva}},
  \bibinfo{author}{\bibfnamefont{M.}~\bibnamefont{Farle}},
  \bibinfo{author}{\bibfnamefont{G.}~\bibnamefont{Dumpich}},
  \bibinfo{author}{\bibfnamefont{H.~Q.} \bibnamefont{Ye}},
  \bibinfo{author}{\bibfnamefont{H.}~\bibnamefont{Poppa}},
  \bibinfo{author}{\bibfnamefont{R.}~\bibnamefont{Kilaas}}, \bibnamefont{and}
  \bibinfo{author}{\bibfnamefont{C.}~\bibnamefont{Kisielowski}},
  \bibinfo{journal}{Phys. Rev. Lett.} \textbf{\bibinfo{volume}{100}},
  \bibinfo{pages}{017205} (\bibinfo{year}{2008}).

\bibitem[{\citenamefont{Stappert et~al.}(2003)\citenamefont{Stappert,
  Rellinghaus, Acet, and Wassermann}}]{Wmann:03}
\bibinfo{author}{\bibfnamefont{S.}~\bibnamefont{Stappert}},
  \bibinfo{author}{\bibfnamefont{B.}~\bibnamefont{Rellinghaus}},
  \bibinfo{author}{\bibfnamefont{M.}~\bibnamefont{Acet}}, \bibnamefont{and}
  \bibinfo{author}{\bibfnamefont{E.~F.} \bibnamefont{Wassermann}},
  \bibinfo{journal}{J. Cryst. Growth} \textbf{\bibinfo{volume}{252}},
  \bibinfo{pages}{440} (\bibinfo{year}{2003}).

\bibitem[{\citenamefont{Rellinghaus et~al.}(2003)\citenamefont{Rellinghaus,
  Stappert, Acet, and Wassermann}}]{WassMannFePt:03}
\bibinfo{author}{\bibfnamefont{B.}~\bibnamefont{Rellinghaus}},
  \bibinfo{author}{\bibfnamefont{S.}~\bibnamefont{Stappert}},
  \bibinfo{author}{\bibfnamefont{M.}~\bibnamefont{Acet}}, \bibnamefont{and}
  \bibinfo{author}{\bibfnamefont{E.~F.} \bibnamefont{Wassermann}},
  \bibinfo{journal}{J. Magn. Magn. Mater.} \textbf{\bibinfo{volume}{266}},
  \bibinfo{pages}{142} (\bibinfo{year}{2003}).

\bibitem[{\citenamefont{Dmitrieva et~al.}(2005)\citenamefont{Dmitrieva,
  Rellinghaus, K\"astner, Liedke, and Fassbender}}]{Fassbender:05}
\bibinfo{author}{\bibfnamefont{O.}~\bibnamefont{Dmitrieva}},
  \bibinfo{author}{\bibfnamefont{B.}~\bibnamefont{Rellinghaus}},
  \bibinfo{author}{\bibfnamefont{J.}~\bibnamefont{K\"astner}},
  \bibinfo{author}{\bibfnamefont{M.~O.} \bibnamefont{Liedke}},
  \bibnamefont{and}
  \bibinfo{author}{\bibfnamefont{J.}~\bibnamefont{Fassbender}},
  \bibinfo{journal}{J. Appl. Phys.} \textbf{\bibinfo{volume}{97}},
  \bibinfo{pages}{10N112} (\bibinfo{year}{2005}).

\bibitem[{\citenamefont{Wulff}(1901)}]{Wulff:01}
\bibinfo{author}{\bibfnamefont{G.}~\bibnamefont{Wulff}}, \bibinfo{journal}{Z.
  Kristallogr.} \textbf{\bibinfo{volume}{34}}, \bibinfo{pages}{449}
  (\bibinfo{year}{1901}).

\bibitem[{\citenamefont{Yamashita et~al.}(1997)\citenamefont{Yamashita, Iwata,
  and Tsunashima}}]{Yamashita:97}
\bibinfo{author}{\bibfnamefont{S.}~\bibnamefont{Yamashita}},
  \bibinfo{author}{\bibfnamefont{S.}~\bibnamefont{Iwata}}, \bibnamefont{and}
  \bibinfo{author}{\bibfnamefont{S.}~\bibnamefont{Tsunashima}},
  \bibinfo{journal}{J. Magn. Soc. Jpn.} \textbf{\bibinfo{volume}{21}},
  \bibinfo{pages}{433} (\bibinfo{year}{1997}).

\bibitem[{\citenamefont{Huang et~al.}(1999{\natexlab{a}})\citenamefont{Huang,
  Hsu, and Lee}}]{Huang:99}
\bibinfo{author}{\bibfnamefont{J.~C.~A.} \bibnamefont{Huang}},
  \bibinfo{author}{\bibfnamefont{A.~C.} \bibnamefont{Hsu}}, \bibnamefont{and}
  \bibinfo{author}{\bibfnamefont{Y.~H.} \bibnamefont{Lee}},
  \bibinfo{journal}{J. Appl. Phys.} \textbf{\bibinfo{volume}{85}},
  \bibinfo{pages}{5977} (\bibinfo{year}{1999}{\natexlab{a}}).

\bibitem[{\citenamefont{Clark et~al.}(1995)\citenamefont{Clark, Pinski,
  Johnson, Sterne, Staunton, and Ginatempo}}]{Clark:95}
\bibinfo{author}{\bibfnamefont{J.~F.} \bibnamefont{Clark}},
  \bibinfo{author}{\bibfnamefont{F.~J.} \bibnamefont{Pinski}},
  \bibinfo{author}{\bibfnamefont{D.~D.} \bibnamefont{Johnson}},
  \bibinfo{author}{\bibfnamefont{P.~A.} \bibnamefont{Sterne}},
  \bibinfo{author}{\bibfnamefont{J.~B.} \bibnamefont{Staunton}},
  \bibnamefont{and}
  \bibinfo{author}{\bibfnamefont{B.}~\bibnamefont{Ginatempo}},
  \bibinfo{journal}{Phys. Rev. Lett.} \textbf{\bibinfo{volume}{74}},
  \bibinfo{pages}{3225} (\bibinfo{year}{1995}).

\bibitem[{\citenamefont{Lu et~al.}(1991)\citenamefont{Lu, Wei, Zunger,
  Frota-Pessoa, and Ferreira}}]{Zunger:91}
\bibinfo{author}{\bibfnamefont{Z.~W.} \bibnamefont{Lu}},
  \bibinfo{author}{\bibfnamefont{S.-H.} \bibnamefont{Wei}},
  \bibinfo{author}{\bibfnamefont{A.}~\bibnamefont{Zunger}},
  \bibinfo{author}{\bibfnamefont{S.}~\bibnamefont{Frota-Pessoa}},
  \bibnamefont{and} \bibinfo{author}{\bibfnamefont{L.~G.}
  \bibnamefont{Ferreira}}, \bibinfo{journal}{Phys. Rev. B}
  \textbf{\bibinfo{volume}{44}}, \bibinfo{pages}{512} (\bibinfo{year}{1991}).

\bibitem[{\citenamefont{Takizawa et~al.}(1991)\citenamefont{Takizawa, Bl\"ugel,
  Terakura, and Oguchi}}]{Bluegel:91}
\bibinfo{author}{\bibfnamefont{S.}~\bibnamefont{Takizawa}},
  \bibinfo{author}{\bibfnamefont{S.}~\bibnamefont{Bl\"ugel}},
  \bibinfo{author}{\bibfnamefont{K.}~\bibnamefont{Terakura}}, \bibnamefont{and}
  \bibinfo{author}{\bibfnamefont{T.}~\bibnamefont{Oguchi}},
  \bibinfo{journal}{Phys. Rev. B} \textbf{\bibinfo{volume}{43}},
  \bibinfo{pages}{947} (\bibinfo{year}{1991}).

\bibitem[{\citenamefont{Sato et~al.}(2008)\citenamefont{Sato, Shimatsu,
  Okazaki, Muraoka, Aoi, Okamoto, and Kitakami}}]{Sato:08}
\bibinfo{author}{\bibfnamefont{H.}~\bibnamefont{Sato}},
  \bibinfo{author}{\bibfnamefont{T.}~\bibnamefont{Shimatsu}},
  \bibinfo{author}{\bibfnamefont{Y.}~\bibnamefont{Okazaki}},
  \bibinfo{author}{\bibfnamefont{H.}~\bibnamefont{Muraoka}},
  \bibinfo{author}{\bibfnamefont{H.}~\bibnamefont{Aoi}},
  \bibinfo{author}{\bibfnamefont{S.}~\bibnamefont{Okamoto}}, \bibnamefont{and}
  \bibinfo{author}{\bibfnamefont{O.}~\bibnamefont{Kitakami}},
  \bibinfo{journal}{J. Appl. Phys.} \textbf{\bibinfo{volume}{103}},
  \bibinfo{pages}{07E114} (\bibinfo{year}{2008}).

\bibitem[{\citenamefont{Huang et~al.}(1999{\natexlab{b}})\citenamefont{Huang,
  Wu, Hsu, Wu, and Hu}}]{Hu:99}
\bibinfo{author}{\bibfnamefont{J.~C.~A.} \bibnamefont{Huang}},
  \bibinfo{author}{\bibfnamefont{T.~H.} \bibnamefont{Wu}},
  \bibinfo{author}{\bibfnamefont{A.~C.} \bibnamefont{Hsu}},
  \bibinfo{author}{\bibfnamefont{L.~C.} \bibnamefont{Wu}}, \bibnamefont{and}
  \bibinfo{author}{\bibfnamefont{Y.~M.} \bibnamefont{Hu}}, \bibinfo{journal}{J.
  Magn. Magn. Mater.} \textbf{\bibinfo{volume}{193}}, \bibinfo{pages}{166}
  (\bibinfo{year}{1999}{\natexlab{b}}).

\bibitem[{\citenamefont{B{\l}o\'nski and Kiejna}(2007)}]{Kiejna:07}
\bibinfo{author}{\bibfnamefont{P.}~\bibnamefont{B{\l}o\'nski}}
  \bibnamefont{and} \bibinfo{author}{\bibfnamefont{A.}~\bibnamefont{Kiejna}},
  \bibinfo{journal}{Surf. Sci.} \textbf{\bibinfo{volume}{601}},
  \bibinfo{pages}{123} (\bibinfo{year}{2007}).

\bibitem[{\citenamefont{Wu and Freeman}(1993)}]{Freeman:93}
\bibinfo{author}{\bibfnamefont{R.}~\bibnamefont{Wu}} \bibnamefont{and}
  \bibinfo{author}{\bibfnamefont{A.~J.} \bibnamefont{Freeman}},
  \bibinfo{journal}{Phys. Rev. B} \textbf{\bibinfo{volume}{47}},
  \bibinfo{pages}{3904} (\bibinfo{year}{1993}).

\bibitem[{\citenamefont{Ohnishi et~al.}(1983)\citenamefont{Ohnishi, Freeman,
  and Weinert}}]{Weinert:83}
\bibinfo{author}{\bibfnamefont{S.}~\bibnamefont{Ohnishi}},
  \bibinfo{author}{\bibfnamefont{A.~J.} \bibnamefont{Freeman}},
  \bibnamefont{and} \bibinfo{author}{\bibfnamefont{M.}~\bibnamefont{Weinert}},
  \bibinfo{journal}{Phys. Rev. B} \textbf{\bibinfo{volume}{28}},
  \bibinfo{pages}{6741} (\bibinfo{year}{1983}).

\bibitem[{\citenamefont{Wang and Freeman}(1981)}]{Wang:81}
\bibinfo{author}{\bibfnamefont{C.~S.} \bibnamefont{Wang}} \bibnamefont{and}
  \bibinfo{author}{\bibfnamefont{A.~J.} \bibnamefont{Freeman}},
  \bibinfo{journal}{Phys. Rev. B} \textbf{\bibinfo{volume}{24}},
  \bibinfo{pages}{4364} (\bibinfo{year}{1981}).

\bibitem[{\citenamefont{Methfessel and Fiorentini}(1996)}]{Fiorentini:96}
\bibinfo{author}{\bibfnamefont{M.}~\bibnamefont{Methfessel}} \bibnamefont{and}
  \bibinfo{author}{\bibfnamefont{V.}~\bibnamefont{Fiorentini}},
  \bibinfo{journal}{J. Phys.: Condens. Matter} \textbf{\bibinfo{volume}{8}},
  \bibinfo{pages}{6525} (\bibinfo{year}{1996}).

\bibitem[{\citenamefont{{\rm Da Silva} et~al.}(2006)\citenamefont{{\rm Da
  Silva}, Stampfl, and Scheffler}}]{Scheffler:06}
\bibinfo{author}{\bibfnamefont{J.~L.~F.} \bibnamefont{{\rm Da Silva}}},
  \bibinfo{author}{\bibfnamefont{C.}~\bibnamefont{Stampfl}}, \bibnamefont{and}
  \bibinfo{author}{\bibfnamefont{M.}~\bibnamefont{Scheffler}},
  \bibinfo{journal}{Surf. Sci.} \textbf{\bibinfo{volume}{600}},
  \bibinfo{pages}{703} (\bibinfo{year}{2006}).

\bibitem[{\citenamefont{Methfessel et~al.}(1991)\citenamefont{Methfessel,
  Hennig, and Scheffler}}]{Scheffler:91}
\bibinfo{author}{\bibfnamefont{M.}~\bibnamefont{Methfessel}},
  \bibinfo{author}{\bibfnamefont{D.}~\bibnamefont{Hennig}}, \bibnamefont{and}
  \bibinfo{author}{\bibfnamefont{M.}~\bibnamefont{Scheffler}},
  \bibinfo{journal}{Phys. Rev. B} \textbf{\bibinfo{volume}{46}},
  \bibinfo{pages}{4816} (\bibinfo{year}{1992}).

\bibitem[{\citenamefont{Fiorentini et~al.}(1993)\citenamefont{Fiorentini,
  Methfessel, and Scheffler}}]{Fiorentini:93}
\bibinfo{author}{\bibfnamefont{V.}~\bibnamefont{Fiorentini}},
  \bibinfo{author}{\bibfnamefont{M.}~\bibnamefont{Methfessel}},
  \bibnamefont{and}
  \bibinfo{author}{\bibfnamefont{M.}~\bibnamefont{Scheffler}},
  \bibinfo{journal}{Phys. Rev. Lett.} \textbf{\bibinfo{volume}{71}},
  \bibinfo{pages}{1051} (\bibinfo{year}{1993}).

\bibitem[{\citenamefont{Skriver and Rosengaard}(1992)}]{Rosengaard:92}
\bibinfo{author}{\bibfnamefont{H.~L.} \bibnamefont{Skriver}} \bibnamefont{and}
  \bibinfo{author}{\bibfnamefont{N.~M.} \bibnamefont{Rosengaard}},
  \bibinfo{journal}{Phys. Rev. B} \textbf{\bibinfo{volume}{46}},
  \bibinfo{pages}{7157} (\bibinfo{year}{1992}).

\bibitem[{\citenamefont{Lu et~al.}(2005)\citenamefont{Lu, Huang, Cuma, and
  Liu}}]{Lu:05}
\bibinfo{author}{\bibfnamefont{G.-H.} \bibnamefont{Lu}},
  \bibinfo{author}{\bibfnamefont{M.}~\bibnamefont{Huang}},
  \bibinfo{author}{\bibfnamefont{M.}~\bibnamefont{Cuma}}, \bibnamefont{and}
  \bibinfo{author}{\bibfnamefont{F.}~\bibnamefont{Liu}},
  \bibinfo{journal}{Surf. Sci.} \textbf{\bibinfo{volume}{588}},
  \bibinfo{pages}{61} (\bibinfo{year}{2005}).

\bibitem[{\citenamefont{Kiejna}(2005)}]{Kiejna:05}
\bibinfo{author}{\bibfnamefont{A.}~\bibnamefont{Kiejna}},
  \bibinfo{journal}{Surf. Sci.} \textbf{\bibinfo{volume}{598}},
  \bibinfo{pages}{276} (\bibinfo{year}{2005}).

\bibitem[{\citenamefont{Kiejna et~al.}(1999)\citenamefont{Kiejna, Peisert, and
  Scharoch}}]{Kiejna:99}
\bibinfo{author}{\bibfnamefont{A.}~\bibnamefont{Kiejna}},
  \bibinfo{author}{\bibfnamefont{J.}~\bibnamefont{Peisert}}, \bibnamefont{and}
  \bibinfo{author}{\bibfnamefont{P.}~\bibnamefont{Scharoch}},
  \bibinfo{journal}{Surf. Sci.} \textbf{\bibinfo{volume}{432}},
  \bibinfo{pages}{54} (\bibinfo{year}{1999}).

\bibitem[{\citenamefont{Chen et~al.}(2003)\citenamefont{Chen, Neyman,
  Gordienko, and R\"osch}}]{Roesch:03}
\bibinfo{author}{\bibfnamefont{Z.-X.} \bibnamefont{Chen}},
  \bibinfo{author}{\bibfnamefont{K.~M.} \bibnamefont{Neyman}},
  \bibinfo{author}{\bibfnamefont{A.~B.} \bibnamefont{Gordienko}},
  \bibnamefont{and} \bibinfo{author}{\bibfnamefont{N.}~\bibnamefont{R\"osch}},
  \bibinfo{journal}{Phys. Rev. B} \textbf{\bibinfo{volume}{68}},
  \bibinfo{pages}{075417} (\bibinfo{year}{2003}).

\bibitem[{\citenamefont{Hong and Yoo}(2005)}]{Yoo:05}
\bibinfo{author}{\bibfnamefont{S.}~\bibnamefont{Hong}} \bibnamefont{and}
  \bibinfo{author}{\bibfnamefont{M.~H.} \bibnamefont{Yoo}},
  \bibinfo{journal}{J. Appl. Phys.} \textbf{\bibinfo{volume}{97}},
  \bibinfo{pages}{084315} (\bibinfo{year}{2005}).

\bibitem[{\citenamefont{Kresse and Furthm\"uller}(1996)}]{Furthmueller:96}
\bibinfo{author}{\bibfnamefont{G.}~\bibnamefont{Kresse}} \bibnamefont{and}
  \bibinfo{author}{\bibfnamefont{J.}~\bibnamefont{Furthm\"uller}},
  \bibinfo{journal}{Phys. Rev. B} \textbf{\bibinfo{volume}{54}},
  \bibinfo{pages}{11169} (\bibinfo{year}{1996}).

\bibitem[{\citenamefont{Reuter et~al.}(2005)\citenamefont{Reuter, Stampfl, and
  Scheffler}}]{SchefflerHandbook:05}
\bibinfo{author}{\bibfnamefont{K.}~\bibnamefont{Reuter}},
  \bibinfo{author}{\bibfnamefont{C.}~\bibnamefont{Stampfl}}, \bibnamefont{and}
  \bibinfo{author}{\bibfnamefont{M.}~\bibnamefont{Scheffler}}, in
  \emph{\bibinfo{booktitle}{Handbook of Materials Modeling}}, edited by
  \bibinfo{editor}{\bibfnamefont{S.}~\bibnamefont{Yip}}
  (\bibinfo{publisher}{Springer}, \bibinfo{address}{Berlin Heidelberg},
  \bibinfo{year}{2005}), vol.~\bibinfo{volume}{1}, p. \bibinfo{pages}{149}.

\bibitem[{\citenamefont{Lee et~al.}(2000)\citenamefont{Lee, Moritz, and
  Scheffler}}]{Scheffler:00}
\bibinfo{author}{\bibfnamefont{S.-H.} \bibnamefont{Lee}},
  \bibinfo{author}{\bibfnamefont{W.}~\bibnamefont{Moritz}}, \bibnamefont{and}
  \bibinfo{author}{\bibfnamefont{M.}~\bibnamefont{Scheffler}},
  \bibinfo{journal}{Phys. Rev. Lett.} \textbf{\bibinfo{volume}{85}},
  \bibinfo{pages}{3890} (\bibinfo{year}{2000}).

\bibitem[{\citenamefont{Moll et~al.}(1996)\citenamefont{Moll, Kley, Pehlke, and
  Scheffler}}]{Moll:96}
\bibinfo{author}{\bibfnamefont{N.}~\bibnamefont{Moll}},
  \bibinfo{author}{\bibfnamefont{A.}~\bibnamefont{Kley}},
  \bibinfo{author}{\bibfnamefont{E.}~\bibnamefont{Pehlke}}, \bibnamefont{and}
  \bibinfo{author}{\bibfnamefont{M.}~\bibnamefont{Scheffler}},
  \bibinfo{journal}{Phys. Rev. B} \textbf{\bibinfo{volume}{54}},
  \bibinfo{pages}{8844} (\bibinfo{year}{1996}).

\bibitem[{\citenamefont{Northrup and Froyen}(1993)}]{Northrup:93}
\bibinfo{author}{\bibfnamefont{J.~E.} \bibnamefont{Northrup}} \bibnamefont{and}
  \bibinfo{author}{\bibfnamefont{S.}~\bibnamefont{Froyen}},
  \bibinfo{journal}{Phys. Rev. Lett.} \textbf{\bibinfo{volume}{71}},
  \bibinfo{pages}{2276} (\bibinfo{year}{1993}).

\bibitem[{\citenamefont{Penev and Kratzer}(2005)}]{Penev:05}
\bibinfo{author}{\bibfnamefont{E.}~\bibnamefont{Penev}} \bibnamefont{and}
  \bibinfo{author}{\bibfnamefont{P.}~\bibnamefont{Kratzer}}, in
  \emph{\bibinfo{booktitle}{Quantum Dots: Fundamentals, Applications, and
  Frontiers}}, edited by \bibinfo{editor}{\bibfnamefont{B.~A.}
  \bibnamefont{Joyce}} (\bibinfo{publisher}{Springer}, \bibinfo{address}{The
  Netherlands}, \bibinfo{year}{2005}), vol. \bibinfo{volume}{190},
  p.~\bibinfo{pages}{27}.

\bibitem[{\citenamefont{Kratzer et~al.}(2003)\citenamefont{Kratzer, Penev, and
  Scheffler}}]{PenevScheff:03}
\bibinfo{author}{\bibfnamefont{P.}~\bibnamefont{Kratzer}},
  \bibinfo{author}{\bibfnamefont{E.}~\bibnamefont{Penev}}, \bibnamefont{and}
  \bibinfo{author}{\bibfnamefont{M.}~\bibnamefont{Scheffler}},
  \bibinfo{journal}{Appl. Surf. Sci.} \textbf{\bibinfo{volume}{216}},
  \bibinfo{pages}{436} (\bibinfo{year}{2003}).

\bibitem[{\citenamefont{Kitchin et~al.}(2008)\citenamefont{Kitchin, Reuter, and
  Scheffler}}]{Kitchin:08}
\bibinfo{author}{\bibfnamefont{J.~R.} \bibnamefont{Kitchin}},
  \bibinfo{author}{\bibfnamefont{K.}~\bibnamefont{Reuter}}, \bibnamefont{and}
  \bibinfo{author}{\bibfnamefont{M.}~\bibnamefont{Scheffler}},
  \bibinfo{journal}{Phys. Rev. B} \textbf{\bibinfo{volume}{77}},
  \bibinfo{pages}{075437} (\bibinfo{year}{2008}).

\bibitem[{\citenamefont{Kresse and Joubert}(1999)}]{Kresse:99}
\bibinfo{author}{\bibfnamefont{G.}~\bibnamefont{Kresse}} \bibnamefont{and}
  \bibinfo{author}{\bibfnamefont{D.}~\bibnamefont{Joubert}},
  \bibinfo{journal}{Phys. Rev. B} \textbf{\bibinfo{volume}{59}},
  \bibinfo{pages}{1758} (\bibinfo{year}{1999}).

\bibitem[{\citenamefont{Perdew et~al.}(1996)\citenamefont{Perdew, Burke, and
  Ernzerhof}}]{Ernzerhof:96}
\bibinfo{author}{\bibfnamefont{J.~P.} \bibnamefont{Perdew}},
  \bibinfo{author}{\bibfnamefont{K.}~\bibnamefont{Burke}}, \bibnamefont{and}
  \bibinfo{author}{\bibfnamefont{M.}~\bibnamefont{Ernzerhof}},
  \bibinfo{journal}{Phys. Rev. Lett.} \textbf{\bibinfo{volume}{77}},
  \bibinfo{pages}{3865} (\bibinfo{year}{1996}).

\bibitem[{\citenamefont{Perdew et~al.}(1992)\citenamefont{Perdew, Chevary,
  Vosko, Jackson, Pederson, Singh, and Fiolhais}}]{Perdew:92}
\bibinfo{author}{\bibfnamefont{J.~P.} \bibnamefont{Perdew}},
  \bibinfo{author}{\bibfnamefont{J.~A.} \bibnamefont{Chevary}},
  \bibinfo{author}{\bibfnamefont{S.~H.} \bibnamefont{Vosko}},
  \bibinfo{author}{\bibfnamefont{K.~A.} \bibnamefont{Jackson}},
  \bibinfo{author}{\bibfnamefont{M.~R.} \bibnamefont{Pederson}},
  \bibinfo{author}{\bibfnamefont{D.~J.} \bibnamefont{Singh}}, \bibnamefont{and}
  \bibinfo{author}{\bibfnamefont{C.}~\bibnamefont{Fiolhais}},
  \bibinfo{journal}{Phys. Rev. B} \textbf{\bibinfo{volume}{46}},
  \bibinfo{pages}{6671} (\bibinfo{year}{1992}).

\bibitem[{\citenamefont{Gross}(2003)}]{Gross:03}
\bibinfo{author}{\bibfnamefont{A.}~\bibnamefont{Gross}},
  \emph{\bibinfo{title}{Theoretical Surface Science-A Microscopic Perspective}}
  (\bibinfo{publisher}{Springer, Berlin}, \bibinfo{year}{2003}).

\bibitem[{\citenamefont{Boettger}(1994)}]{Boettger:94}
\bibinfo{author}{\bibfnamefont{J.~C.} \bibnamefont{Boettger}},
  \bibinfo{journal}{Phys. Rev. B} \textbf{\bibinfo{volume}{49}},
  \bibinfo{pages}{16798} (\bibinfo{year}{1994}).

\bibitem[{\citenamefont{Zangwill}(1988)}]{Zangwill:88}
\bibinfo{author}{\bibfnamefont{A.}~\bibnamefont{Zangwill}},
  \emph{\bibinfo{title}{Physics at Surfaces}} (\bibinfo{publisher}{Cambridge
  University Press}, \bibinfo{address}{Cambridge}, \bibinfo{year}{1988}).

\bibitem[{\citenamefont{Penev}(2002)}]{Penev:02}
\bibinfo{author}{\bibfnamefont{E.~S.} \bibnamefont{Penev}}, Ph.D. thesis,
  \bibinfo{school}{Technische Universit\"at Berlin} (\bibinfo{year}{2002}).

\bibitem[{\citenamefont{Iddir et~al.}(2007)\citenamefont{Iddir, Komanicky,
  \"O\v{g}\"ut, You, and Zapol}}]{Iddir:07}
\bibinfo{author}{\bibfnamefont{H.}~\bibnamefont{Iddir}},
  \bibinfo{author}{\bibfnamefont{V.}~\bibnamefont{Komanicky}},
  \bibinfo{author}{\bibnamefont{\"O\v{g}\"ut}},
  \bibinfo{author}{\bibfnamefont{H.}~\bibnamefont{You}}, \bibnamefont{and}
  \bibinfo{author}{\bibfnamefont{P.}~\bibnamefont{Zapol}}, \bibinfo{journal}{J.
  Phys. Chem. C} \textbf{\bibinfo{volume}{111}}, \bibinfo{pages}{14782}
  (\bibinfo{year}{2007}).

\bibitem[{\citenamefont{Getman and Schneider}(2007)}]{Getman:07}
\bibinfo{author}{\bibfnamefont{R.~B.} \bibnamefont{Getman}} \bibnamefont{and}
  \bibinfo{author}{\bibfnamefont{W.~F.} \bibnamefont{Schneider}},
  \bibinfo{journal}{J. Phys. Chem. C} \textbf{\bibinfo{volume}{111}},
  \bibinfo{pages}{389} (\bibinfo{year}{2007}).

\bibitem[{\citenamefont{B{\l}o\'nski and Kiejna}(2004)}]{Blonski:04}
\bibinfo{author}{\bibfnamefont{P.}~\bibnamefont{B{\l}o\'nski}}
  \bibnamefont{and} \bibinfo{author}{\bibfnamefont{A.}~\bibnamefont{Kiejna}},
  \bibinfo{journal}{Vacuum} \textbf{\bibinfo{volume}{74}}, \bibinfo{pages}{179}
  (\bibinfo{year}{2004}).

\bibitem[{\citenamefont{Spencer et~al.}(2002)\citenamefont{Spencer, Hung,
  Snook, and Yarovsky}}]{Spencer:02}
\bibinfo{author}{\bibfnamefont{M.~J.~S.} \bibnamefont{Spencer}},
  \bibinfo{author}{\bibfnamefont{A.}~\bibnamefont{Hung}},
  \bibinfo{author}{\bibfnamefont{I.~K.} \bibnamefont{Snook}}, \bibnamefont{and}
  \bibinfo{author}{\bibfnamefont{I.}~\bibnamefont{Yarovsky}},
  \bibinfo{journal}{Surf. Sci.} \textbf{\bibinfo{volume}{513}},
  \bibinfo{pages}{389} (\bibinfo{year}{2002}).

\bibitem[{\citenamefont{Vitos et~al.}(1998)\citenamefont{Vitos, Ruban, Skriver,
  and Koll\'ar}}]{Vitos:98}
\bibinfo{author}{\bibfnamefont{L.}~\bibnamefont{Vitos}},
  \bibinfo{author}{\bibfnamefont{A.~V.} \bibnamefont{Ruban}},
  \bibinfo{author}{\bibfnamefont{H.~L.} \bibnamefont{Skriver}},
  \bibnamefont{and} \bibinfo{author}{\bibfnamefont{J.}~\bibnamefont{Koll\'ar}},
  \bibinfo{journal}{Surf. Sci.} \textbf{\bibinfo{volume}{411}},
  \bibinfo{pages}{186} (\bibinfo{year}{1998}).

\bibitem[{\citenamefont{Ald\'en et~al.}(1992)\citenamefont{Ald\'en, Mirbt,
  Skriver, Rosengaard, and Johansson}}]{Alden:92}
\bibinfo{author}{\bibfnamefont{M.}~\bibnamefont{Ald\'en}},
  \bibinfo{author}{\bibfnamefont{S.}~\bibnamefont{Mirbt}},
  \bibinfo{author}{\bibfnamefont{H.~L.} \bibnamefont{Skriver}},
  \bibinfo{author}{\bibfnamefont{N.~M.} \bibnamefont{Rosengaard}},
  \bibnamefont{and}
  \bibinfo{author}{\bibfnamefont{B.}~\bibnamefont{Johansson}},
  \bibinfo{journal}{Phys. Rev. B} \textbf{\bibinfo{volume}{46}},
  \bibinfo{pages}{6303} (\bibinfo{year}{1992}).

\bibitem[{\citenamefont{Mehl and
  Papaconstantopoulos}(1996)}]{Papaconstantopoulos:96}
\bibinfo{author}{\bibfnamefont{M.~J.} \bibnamefont{Mehl}} \bibnamefont{and}
  \bibinfo{author}{\bibfnamefont{D.~A.} \bibnamefont{Papaconstantopoulos}},
  \bibinfo{journal}{Phys. Rev. B} \textbf{\bibinfo{volume}{54}},
  \bibinfo{pages}{4519} (\bibinfo{year}{1996}).

\bibitem[{\citenamefont{Foiles et~al.}(1986)\citenamefont{Foiles, Baskes, and
  Daw}}]{Baskes86:86}
\bibinfo{author}{\bibfnamefont{S.~M.} \bibnamefont{Foiles}},
  \bibinfo{author}{\bibfnamefont{M.~I.} \bibnamefont{Baskes}},
  \bibnamefont{and} \bibinfo{author}{\bibfnamefont{M.~S.} \bibnamefont{Daw}},
  \bibinfo{journal}{Phys. Rev. B} \textbf{\bibinfo{volume}{33}},
  \bibinfo{pages}{7983} (\bibinfo{year}{1986}).

\bibitem[{\citenamefont{Baskes}(1992)}]{Baskes:92}
\bibinfo{author}{\bibfnamefont{M.~I.} \bibnamefont{Baskes}},
  \bibinfo{journal}{Phys. Rev. B} \textbf{\bibinfo{volume}{46}},
  \bibinfo{pages}{2727} (\bibinfo{year}{1992}).

\bibitem[{\citenamefont{Tyson and Miller}(1977)}]{Miller:77}
\bibinfo{author}{\bibfnamefont{W.~R.} \bibnamefont{Tyson}} \bibnamefont{and}
  \bibinfo{author}{\bibfnamefont{W.~A.} \bibnamefont{Miller}},
  \bibinfo{journal}{Surf. Sci.} \textbf{\bibinfo{volume}{62}},
  \bibinfo{pages}{267} (\bibinfo{year}{1977}).

\bibitem[{\citenamefont{de~Boer et~al.}(1989)\citenamefont{de~Boer, Boom,
  Mattens, Miedema, and Niessen}}]{Boer:89}
\bibinfo{author}{\bibfnamefont{F.~R.} \bibnamefont{de~Boer}},
  \bibinfo{author}{\bibfnamefont{R.}~\bibnamefont{Boom}},
  \bibinfo{author}{\bibfnamefont{W.~C.~M.} \bibnamefont{Mattens}},
  \bibinfo{author}{\bibfnamefont{A.~R.} \bibnamefont{Miedema}},
  \bibnamefont{and} \bibinfo{author}{\bibfnamefont{A.~K.}
  \bibnamefont{Niessen}}, \emph{\bibinfo{title}{\rm Cohesion in metals}}
  (\bibinfo{publisher}{North-Holland Physics Publishing},
  \bibinfo{year}{1989}).

\bibitem[{\citenamefont{Tyson}(1975)}]{Tyson:75}
\bibinfo{author}{\bibfnamefont{W.~R.} \bibnamefont{Tyson}},
  \bibinfo{journal}{Can. Met. Quart.} \textbf{\bibinfo{volume}{14}},
  \bibinfo{pages}{307} (\bibinfo{year}{1975}).

\bibitem[{\citenamefont{Herper et~al.}(1999)\citenamefont{Herper, Hoffmann, and
  Entel}}]{Herper:99}
\bibinfo{author}{\bibfnamefont{H.~C.} \bibnamefont{Herper}},
  \bibinfo{author}{\bibfnamefont{E.}~\bibnamefont{Hoffmann}}, \bibnamefont{and}
  \bibinfo{author}{\bibfnamefont{P.}~\bibnamefont{Entel}},
  \bibinfo{journal}{Phys. Rev. B} \textbf{\bibinfo{volume}{60}},
  \bibinfo{pages}{3839} (\bibinfo{year}{1999}).

\bibitem[{\citenamefont{Entel et~al.}(2000)\citenamefont{Entel, Herper,
  Hoffmann, Nepecks, Wassermann, Acet, Crisan, and Akai}}]{EntelIron:00}
\bibinfo{author}{\bibfnamefont{P.}~\bibnamefont{Entel}},
  \bibinfo{author}{\bibfnamefont{H.~C.} \bibnamefont{Herper}},
  \bibinfo{author}{\bibfnamefont{E.}~\bibnamefont{Hoffmann}},
  \bibinfo{author}{\bibfnamefont{G.}~\bibnamefont{Nepecks}},
  \bibinfo{author}{\bibfnamefont{E.~F.} \bibnamefont{Wassermann}},
  \bibinfo{author}{\bibfnamefont{M.}~\bibnamefont{Acet}},
  \bibinfo{author}{\bibfnamefont{V.}~\bibnamefont{Crisan}}, \bibnamefont{and}
  \bibinfo{author}{\bibfnamefont{H.}~\bibnamefont{Akai}},
  \bibinfo{journal}{Phil. Mag. B} \textbf{\bibinfo{volume}{80}},
  \bibinfo{pages}{141} (\bibinfo{year}{2000}).

\bibitem[{\citenamefont{Acet et~al.}(1994)\citenamefont{Acet, Z\"ahres,
  Wassermann, and Pepperhoff}}]{WassermannIron:94}
\bibinfo{author}{\bibfnamefont{M.}~\bibnamefont{Acet}},
  \bibinfo{author}{\bibfnamefont{H.}~\bibnamefont{Z\"ahres}},
  \bibinfo{author}{\bibfnamefont{E.~F.} \bibnamefont{Wassermann}},
  \bibnamefont{and}
  \bibinfo{author}{\bibfnamefont{W.}~\bibnamefont{Pepperhoff}},
  \bibinfo{journal}{Phys. Rev. B} \textbf{\bibinfo{volume}{49}},
  \bibinfo{pages}{6012} (\bibinfo{year}{1994}).

\bibitem[{\citenamefont{Cleri and Rosato}(1993)}]{Rosato:93}
\bibinfo{author}{\bibfnamefont{F.}~\bibnamefont{Cleri}} \bibnamefont{and}
  \bibinfo{author}{\bibfnamefont{V.}~\bibnamefont{Rosato}},
  \bibinfo{journal}{Phys. Rev. B} \textbf{\bibinfo{volume}{48}},
  \bibinfo{pages}{22} (\bibinfo{year}{1993}).

\bibitem[{\citenamefont{Khein et~al.}(1995)\citenamefont{Khein, Singh, and
  Umrigar}}]{Khein:95}
\bibinfo{author}{\bibfnamefont{A.}~\bibnamefont{Khein}},
  \bibinfo{author}{\bibfnamefont{D.~J.} \bibnamefont{Singh}}, \bibnamefont{and}
  \bibinfo{author}{\bibfnamefont{C.~J.} \bibnamefont{Umrigar}},
  \bibinfo{journal}{Phys. Rev. B} \textbf{\bibinfo{volume}{51}},
  \bibinfo{pages}{4105} (\bibinfo{year}{1995}).

\bibitem[{\citenamefont{Moroni et~al.}(1997)\citenamefont{Moroni, Kresse,
  Hafner, and Furthm\"uller}}]{Kresse:97}
\bibinfo{author}{\bibfnamefont{E.~G.} \bibnamefont{Moroni}},
  \bibinfo{author}{\bibfnamefont{G.}~\bibnamefont{Kresse}},
  \bibinfo{author}{\bibfnamefont{J.}~\bibnamefont{Hafner}}, \bibnamefont{and}
  \bibinfo{author}{\bibfnamefont{J.}~\bibnamefont{Furthm\"uller}},
  \bibinfo{journal}{Phys. Rev. B} \textbf{\bibinfo{volume}{56}},
  \bibinfo{pages}{15629} (\bibinfo{year}{1997}).

\bibitem[{\citenamefont{Kokalj and Caus\`a}(1999)}]{Ptbulk:99}
\bibinfo{author}{\bibfnamefont{A.}~\bibnamefont{Kokalj}} \bibnamefont{and}
  \bibinfo{author}{\bibfnamefont{M.}~\bibnamefont{Caus\`a}},
  \bibinfo{journal}{J. Phys.: Condens. Matter} \textbf{\bibinfo{volume}{11}},
  \bibinfo{pages}{7463} (\bibinfo{year}{1999}).

\bibitem[{\citenamefont{Fox and Jansen}(1999)}]{Cobulk:99}
\bibinfo{author}{\bibfnamefont{S.}~\bibnamefont{Fox}} \bibnamefont{and}
  \bibinfo{author}{\bibfnamefont{H.~J.~F.} \bibnamefont{Jansen}},
  \bibinfo{journal}{Phys. Rev. B} \textbf{\bibinfo{volume}{60}},
  \bibinfo{pages}{4397} (\bibinfo{year}{1999}).

\bibitem[{\citenamefont{Rollmann et~al.}(2004)\citenamefont{Rollmann, Sahoo,
  and Entel}}]{Sahoo:04}
\bibinfo{author}{\bibfnamefont{G.}~\bibnamefont{Rollmann}},
  \bibinfo{author}{\bibfnamefont{S.}~\bibnamefont{Sahoo}}, \bibnamefont{and}
  \bibinfo{author}{\bibfnamefont{P.}~\bibnamefont{Entel}},
  \bibinfo{journal}{Phys. Status Solidi} \textbf{\bibinfo{volume}{201}},
  \bibinfo{pages}{3263} (\bibinfo{year}{2004}).

\bibitem[{\citenamefont{Rollmann et~al.}(2006)\citenamefont{Rollmann, Entel,
  and Sahoo}}]{Sahoo:06}
\bibinfo{author}{\bibfnamefont{G.}~\bibnamefont{Rollmann}},
  \bibinfo{author}{\bibfnamefont{P.}~\bibnamefont{Entel}}, \bibnamefont{and}
  \bibinfo{author}{\bibfnamefont{S.}~\bibnamefont{Sahoo}},
  \bibinfo{journal}{Comp. Mater. Sci.} \textbf{\bibinfo{volume}{35}},
  \bibinfo{pages}{275} (\bibinfo{year}{2006}).

\bibitem[{\citenamefont{Burkert et~al.}(2005)\citenamefont{Burkert, Eriksson,
  Simak, Ruban, Sanyal, Nordstr\"om, and Wills}}]{Wills:05}
\bibinfo{author}{\bibfnamefont{T.}~\bibnamefont{Burkert}},
  \bibinfo{author}{\bibfnamefont{O.}~\bibnamefont{Eriksson}},
  \bibinfo{author}{\bibfnamefont{S.~I.} \bibnamefont{Simak}},
  \bibinfo{author}{\bibfnamefont{A.~V.} \bibnamefont{Ruban}},
  \bibinfo{author}{\bibfnamefont{B.}~\bibnamefont{Sanyal}},
  \bibinfo{author}{\bibfnamefont{L.}~\bibnamefont{Nordstr\"om}},
  \bibnamefont{and} \bibinfo{author}{\bibfnamefont{J.~M.} \bibnamefont{Wills}},
  \bibinfo{journal}{Phys. Rev. B} \textbf{\bibinfo{volume}{71}},
  \bibinfo{pages}{134411} (\bibinfo{year}{2005}).

\bibitem[{\citenamefont{Ravindran et~al.}(2001)\citenamefont{Ravindran,
  Kjekshus, Fjellvag, James, Nordstr\"om, Johansson, and
  Eriksson}}]{Eriksson:01}
\bibinfo{author}{\bibfnamefont{P.}~\bibnamefont{Ravindran}},
  \bibinfo{author}{\bibfnamefont{A.}~\bibnamefont{Kjekshus}},
  \bibinfo{author}{\bibfnamefont{H.}~\bibnamefont{Fjellvaag}},
  \bibinfo{author}{\bibfnamefont{P.}~\bibnamefont{James}},
  \bibinfo{author}{\bibfnamefont{L.}~\bibnamefont{Nordstr\"om}},
  \bibinfo{author}{\bibfnamefont{B.}~\bibnamefont{Johansson}},
  \bibnamefont{and} \bibinfo{author}{\bibfnamefont{O.}~\bibnamefont{Eriksson}},
  \bibinfo{journal}{Phys. Rev. B} \textbf{\bibinfo{volume}{63}},
  \bibinfo{pages}{144409} (\bibinfo{year}{2001}).

\bibitem[{\citenamefont{Galanakis et~al.}(2000)\citenamefont{Galanakis,
  Alouani, and Dreysse}}]{Galanakis:00}
\bibinfo{author}{\bibfnamefont{I.}~\bibnamefont{Galanakis}},
  \bibinfo{author}{\bibfnamefont{M.}~\bibnamefont{Alouani}}, \bibnamefont{and}
  \bibinfo{author}{\bibfnamefont{H.}~\bibnamefont{Dreysse}},
  \bibinfo{journal}{Phys. Rev. B} \textbf{\bibinfo{volume}{62}},
  \bibinfo{pages}{6475} (\bibinfo{year}{2000}).

\bibitem[{\citenamefont{Kashyap et~al.}(1999)\citenamefont{Kashyap, Garg,
  Solanki, Nautiyal, and Auluck}}]{Garg:99}
\bibinfo{author}{\bibfnamefont{A.}~\bibnamefont{Kashyap}},
  \bibinfo{author}{\bibfnamefont{K.~B.} \bibnamefont{Garg}},
  \bibinfo{author}{\bibfnamefont{A.~K.} \bibnamefont{Solanki}},
  \bibinfo{author}{\bibfnamefont{T.}~\bibnamefont{Nautiyal}}, \bibnamefont{and}
  \bibinfo{author}{\bibfnamefont{S.}~\bibnamefont{Auluck}},
  \bibinfo{journal}{Phys. Rev. B} \textbf{\bibinfo{volume}{60}},
  \bibinfo{pages}{2262} (\bibinfo{year}{1999}).

\bibitem[{\citenamefont{Zeng et~al.}(2002)\citenamefont{Zeng, Sabirianov,
  Mryasov, Yan, Cho, and Sellmyer}}]{Zeng:02}
\bibinfo{author}{\bibfnamefont{H.}~\bibnamefont{Zeng}},
  \bibinfo{author}{\bibfnamefont{R.}~\bibnamefont{Sabirianov}},
  \bibinfo{author}{\bibfnamefont{O.}~\bibnamefont{Mryasov}},
  \bibinfo{author}{\bibfnamefont{M.~L.} \bibnamefont{Yan}},
  \bibinfo{author}{\bibfnamefont{K.}~\bibnamefont{Cho}}, \bibnamefont{and}
  \bibinfo{author}{\bibfnamefont{D.~J.} \bibnamefont{Sellmyer}},
  \bibinfo{journal}{Phys. Rev. B} \textbf{\bibinfo{volume}{66}},
  \bibinfo{pages}{184425} (\bibinfo{year}{2002}).

\bibitem[{\citenamefont{Brown et~al.}(2003)\citenamefont{Brown, Kraczek,
  Janotti, Schulthess, Stocks, and Johnson}}]{Brown:03}
\bibinfo{author}{\bibfnamefont{G.}~\bibnamefont{Brown}},
  \bibinfo{author}{\bibfnamefont{B.}~\bibnamefont{Kraczek}},
  \bibinfo{author}{\bibfnamefont{A.}~\bibnamefont{Janotti}},
  \bibinfo{author}{\bibfnamefont{T.~C.} \bibnamefont{Schulthess}},
  \bibinfo{author}{\bibfnamefont{G.~M.} \bibnamefont{Stocks}},
  \bibnamefont{and} \bibinfo{author}{\bibfnamefont{D.~D.}
  \bibnamefont{Johnson}}, \bibinfo{journal}{Phys. Rev. B}
  \textbf{\bibinfo{volume}{68}}, \bibinfo{pages}{052405}
  (\bibinfo{year}{2003}).

\bibitem[{\citenamefont{Dannenberg et~al.}(2009)\citenamefont{Dannenberg,
  Gruner, and Entel}}]{MyICML11:09}
\bibinfo{author}{\bibfnamefont{A.}~\bibnamefont{Dannenberg}},
  \bibinfo{author}{\bibfnamefont{M.~E.} \bibnamefont{Gruner}},
  \bibnamefont{and} \bibinfo{author}{\bibfnamefont{P.}~\bibnamefont{Entel}},
  \bibinfo{journal}{J. Phys.: Conf. Ser.}  (\bibinfo{year}{2009}),
  \bibinfo{note}{to be published}.

\bibitem[{\citenamefont{Smoluchowski}(1941)}]{Smoluchowski:41}
\bibinfo{author}{\bibfnamefont{R.}~\bibnamefont{Smoluchowski}},
  \bibinfo{journal}{Phys. Rev.} \textbf{\bibinfo{volume}{60}},
  \bibinfo{pages}{661} (\bibinfo{year}{1941}).

\bibitem[{\citenamefont{Finnis and Heine}(1974)}]{Finnis:74}
\bibinfo{author}{\bibfnamefont{M.~W.} \bibnamefont{Finnis}} \bibnamefont{and}
  \bibinfo{author}{\bibfnamefont{V.}~\bibnamefont{Heine}}, \bibinfo{journal}{J.
  Phys. F: Metal Phys.} \textbf{\bibinfo{volume}{4}}, \bibinfo{pages}{L37}
  (\bibinfo{year}{1974}).

\bibitem[{\citenamefont{Pettifor}(1978)}]{Pettifor:78}
\bibinfo{author}{\bibfnamefont{D.~G.} \bibnamefont{Pettifor}},
  \bibinfo{journal}{J. Phys. F: Metal Phys.} \textbf{\bibinfo{volume}{8}},
  \bibinfo{pages}{219} (\bibinfo{year}{1978}).

\bibitem[{\citenamefont{Landman and Hill}(1980)}]{Hill:80}
\bibinfo{author}{\bibfnamefont{U.}~\bibnamefont{Landman}}, 
  \bibinfo{author}{\bibfnamefont{R.~N.}~\bibnamefont{Hill}} \bibnamefont{and}
  \bibinfo{author}{\bibfnamefont{M.}~\bibnamefont{Mostoller}}
  \bibinfo{journal}{Phys. Rev. B} \textbf{\bibinfo{volume}{21}},
  \bibinfo{pages}{448} (\bibinfo{year}{1980}).

\bibitem[{\citenamefont{Heine and Marks}(1986)}]{Heine:86}
\bibinfo{author}{\bibfnamefont{V.}~\bibnamefont{Heine}} \bibnamefont{and}
  \bibinfo{author}{\bibfnamefont{L.~D.} \bibnamefont{Marks}},
  \bibinfo{journal}{Surf. Sci.} \textbf{\bibinfo{volume}{165}},
  \bibinfo{pages}{65} (\bibinfo{year}{1986}).

\bibitem[{\citenamefont{Z\'olyomi et~al.}(2009)\citenamefont{Z\'olyomi, Vitos,
  Kwon, and Koll\'ar}}]{Zolyomi:09}
\bibinfo{author}{\bibfnamefont{V.}~\bibnamefont{Z\'olyomi}},
  \bibinfo{author}{\bibfnamefont{L.}~\bibnamefont{Vitos}},
  \bibinfo{author}{\bibfnamefont{S.~K.} \bibnamefont{Kwon}}, \bibnamefont{and}
  \bibinfo{author}{\bibfnamefont{J.}~\bibnamefont{Koll\'ar}},
  \bibinfo{journal}{J. Phys.: Condens. Matter} \textbf{\bibinfo{volume}{21}},
  \bibinfo{pages}{095007} (\bibinfo{year}{2009}).

\bibitem[{\citenamefont{Gruner et~al.}(2008)\citenamefont{Gruner, Rollmann,
  Entel, and Farle}}]{GrunerPartikel:08}
\bibinfo{author}{\bibfnamefont{M.~E.} \bibnamefont{Gruner}},
  \bibinfo{author}{\bibfnamefont{G.}~\bibnamefont{Rollmann}},
  \bibinfo{author}{\bibfnamefont{P.}~\bibnamefont{Entel}}, \bibnamefont{and}
  \bibinfo{author}{\bibfnamefont{M.}~\bibnamefont{Farle}},
  \bibinfo{journal}{Phys. Rev. Lett.} \textbf{\bibinfo{volume}{100}},
  \bibinfo{pages}{087203} (\bibinfo{year}{2008}).

\bibitem[{\citenamefont{Ma and Balbuena}(2008)}]{Ptseg:08}
\bibinfo{author}{\bibfnamefont{Y.}~\bibnamefont{Ma}} \bibnamefont{and}
  \bibinfo{author}{\bibfnamefont{P.~B.} \bibnamefont{Balbuena}},
  \bibinfo{journal}{Surf. Sci.} \textbf{\bibinfo{volume}{602}},
  \bibinfo{pages}{107} (\bibinfo{year}{2008}).

\bibitem[{\citenamefont{Yuge et~al.}(2007)\citenamefont{Yuge, Seko, Kuwabara,
  Oba, and Tanaka}}]{PtsegCuPt:07}
\bibinfo{author}{\bibfnamefont{K.}~\bibnamefont{Yuge}},
  \bibinfo{author}{\bibfnamefont{A.}~\bibnamefont{Seko}},
  \bibinfo{author}{\bibfnamefont{A.}~\bibnamefont{Kuwabara}},
  \bibinfo{author}{\bibfnamefont{F.}~\bibnamefont{Oba}}, \bibnamefont{and}
  \bibinfo{author}{\bibfnamefont{I.}~\bibnamefont{Tanaka}},
  \bibinfo{journal}{Phys. Rev. B} \textbf{\bibinfo{volume}{76}},
  \bibinfo{pages}{045407} (\bibinfo{year}{2007}).

\bibitem[{\citenamefont{Entel and Gruner}(2009)}]{GrunerLatest:08}
\bibinfo{author}{\bibfnamefont{P.}~\bibnamefont{Entel}} \bibnamefont{and}
  \bibinfo{author}{\bibfnamefont{M.~E.} \bibnamefont{Gruner}},
  \bibinfo{journal}{J. Phys.: Condens. Matter} \textbf{\bibinfo{volume}{21}},
  \bibinfo{pages}{064228} (\bibinfo{year}{2009}).

\bibitem[{\citenamefont{Gruner and Entel}(2009)}]{GrunerSub:09}
\bibinfo{author}{\bibfnamefont{M.~E.} \bibnamefont{Gruner}} \bibnamefont{and}
  \bibinfo{author}{\bibfnamefont{P.}~\bibnamefont{Entel}}, \bibinfo{journal}{J.
  Phys.: Condens. Matter} \textbf{\bibinfo{volume}{21}},
  \bibinfo{pages}{293201} (\bibinfo{year}{2009}).

\bibitem[{\citenamefont{Edelstein and Cammarat}(1997)}]{Edelstein:97}
\bibinfo{author}{\bibfnamefont{A.~S.} \bibnamefont{Edelstein}}
  \bibnamefont{and} \bibinfo{author}{\bibfnamefont{R.~C.}
  \bibnamefont{Cammarat}}, \emph{\bibinfo{title}{Nanomaterials: synthesis,
  properties and applications}} (\bibinfo{publisher}{Institute of Physics
  Publishing}, \bibinfo{address}{Bristol}, \bibinfo{year}{1997}).

\bibitem[{\citenamefont{M\"uller and Albe}(2007)}]{Albe:07}
\bibinfo{author}{\bibfnamefont{M.}~\bibnamefont{M\"uller}} \bibnamefont{and}
  \bibinfo{author}{\bibfnamefont{K.}~\bibnamefont{Albe}},
  \bibinfo{journal}{Acta Mater.} \textbf{\bibinfo{volume}{55}},
  \bibinfo{pages}{6617} (\bibinfo{year}{2007}).

\end{thebibliography}
\end{document}